\documentclass[useAMS,usenatbib,fleqn]{mn2e}
\usepackage{txfonts}
\usepackage[pdftex]{graphicx}
\usepackage{graphicx}
\usepackage{natbib}
\usepackage{pdfpages}
\usepackage{subfigure}
\def\mpc{h^{-1} {\rm{Mpc}}}
\def\kms {\rm{km~s^{-1}}}
\def\apj {ApJ}
\def\apjl {ApJL}
\def\apjs {ApJS}
\def\aj {AJ}
\def\aap {A\&A}
\def\mnras {MNRAS}
\def\araa {ARA\&A}
\def\nat {Nature}
\def\arcsec{''}

\def\Msol {\mathcal{M}_\odot}

\def\Gyr {\rm Gyr}
\def\Myr {\rm Myr}

\begin{document}
\title[Green Valley galaxies]
{Green Valley galaxies as a transition population in different environments}
\author[V. Coenda et al.]
{Valeria Coenda\thanks{E-mail:vcoenda@oac.unc.edu.ar}, H\'ector J. Mart\'inez 
and Hern\'an Muriel\\
Instituto de Astronom\'{\i}a Te\'orica y Experimental (IATE), 
CONICET$-$Universidad Nacional de 
C\'ordoba, Laprida 854, X5000BGR, C\'ordoba, Argentina\\
Observatorio Astron\'omico, Universidad Nacional de 
C\'ordoba, Laprida 854, X5000BGR, C\'ordoba, Argentina}
\date{\today}
\pagerange{\pageref{firstpage}--\pageref{lastpage}} 
\maketitle
\label{firstpage}


\begin{abstract}
We present a comparative analysis of the properties of passive, star-forming 
and transition (green valley) galaxies, in four discrete 
environments: field, groups, the outskirts and the core of X-ray clusters. 
We construct samples of galaxies from the SDSS in these environments so that they 
are bound to have similar redshift distributions. The classification
of galaxies into the three sequences is based on the UV-optical colour $NUV-r$. 
We study a number of galaxy properties: stellar mass, 
morphology, specific star formation rate and the history of star formation.
The analysis of green valley galaxies reveals that the physical mechanisms
responsible for external quenching become more efficient moving from the 
field to denser environments.
We confirm previous findings that green valley galaxies have intermediate
morphologies, moreover, we find that this appears to be independent of the 
environment. Regarding the stellar mass of green valley galaxies, we find that 
they tend to be more massive in the field than in denser environments.
On average, green valley galaxies account for $\sim 20\%$ of all galaxies in groups 
and X-ray clusters. We find evidence that the field environment is inefficient 
in transforming low mass galaxies. Green valley galaxies have average star 
formation histories intermediate between passive and star forming galaxies, 
and have a clear and consistent dependence on the environment: both, the quenching 
time, and the amplitude of the star formation rate, decrease towards
higher density environments.
\end{abstract}


\section{Introduction} \label{sec:intro}
Galaxies in the local Universe can be classified into two broad categories: 
they are either passively evolving red galaxies with old stellar populations 
commonly found in high density regions; or blue star-forming galaxies 
preferentially inhabiting
low density regions. The colour bimodality has been observed from low redshift, 
$z\sim 0.1$ (e.g. \citealt{Strateva:2001,Baldry:2004}), up to $z\sim1$ 
(e.g. \citealt{Balogh98, Im:2002,Bell:2004,Weiner:2005,Willmer:2006}).
These two populations of galaxies are usually referred to as the 
\textit{red sequence} (RS) and the \textit{blue cloud} (BC), respectively, in 
the optical colour-magnitude diagram. This bimodality in galaxy colour 
and its correlation with galaxy morphology was first noted by 
\citet{Takamiya:1995}. The advent of the Galaxy Zoo project  
\citep{Lintott:2008} helped to confirm that this bimodality is not entirely 
morphologically driven. For instance, \citet{Schawinski:2009} found a large 
fraction of ellipticals in the BC, and \citet{Masters:2010} found spirals
in the RS.

The origin of the bimodality observed could be due to secular evolution 
(nature scenario), or to the action of environmental physical 
processes (nurture scenario). The secular evolution of a galaxy is much more 
than the mere ageing of its stellar population (\citealt{Masters:2010, 
Masters:2011}, \citealt{Mendez:2011}). Internal processes of 
secular evolution in disk galaxies could have important consequences in the 
process of star formation (i.e. pseudo-bulges, \citealt{KK:2004}). These are slow 
processes with timescales much longer than the dynamic time of a galaxy. The 
evolution of the stellar mass function of star-forming and quiescent galaxies
suggests that galaxies that reach a stellar mass of $\sim ∼ 10^{10.8}M_{\odot}$
quench and become quiescent galaxies (e.g. Peng et al. 2010). This effect is 
known as mass quenching. However, the causes of this decline in star formation are not entirely clear. Among the suggested processes are halo heating
\citep{Marasco:2012} and AGN feedback. 
Some authors propose AGN feedback as the primary mechanism that originates both
the suppression or quenching of star formation in massive galaxies, and the 
correlation of the central black hole mass with the galactic bulge mass 
(e.g. \citealt{DiMatteo:2005, Martin:2007,Schawinski:2007,McConnell:2013}). 
We will refer to the many internal processes that can quench star formation and
are supposed to be independent of the environment, as internal quenching. 
On the other hand, many quenching mechanisms associated with the environment, 
which we will refer to as external quenching, have been proposed. 
One of these mechanisms is ram pressure stripping of the cold gas in galaxies. 
This process has a timescale of a few hundred Myr 
(e.g. \citealt{GG:1972, Abadi:1999}). 
Galaxy-group interactions such as strangulation can remove 
warm and hot gas from a galactic halo, cutting off the supply of gas for star 
formation \citep{Larson:1980, Kawata:2008}. 
On the other hand, halos can play a role in diminishing star 
formation. In massive halos, shock heating could prevent accretion
of cold gas into galaxies, thus inhibiting star formation.  
Galaxy-galaxy interactions in high mass systems are frequent
and produce morphological transformations \citep{Moore:1996, Moore:1999}.

The fraction of red galaxies increases with time (\citealt{Faber:2007}) and, 
therefore, galaxies must transit from blue to red. These transition galaxies
between the RS and the BC are called \textit{green galaxies}, and they inhabit
the so called \textit{green valley} (GV) in the colour-magnitude diagram. 
Several authors have 
defined the GV using optical colours, preferentially $g-r$, or $u-r$, 
however, \citet{Wyder:2007} found that the UV-optical $NUV-r$ colour provides
a more efficient criterion to select GV galaxies. Ultraviolet emission is 
sensitive to recent star formation, while the $r$-band is more sensitive to the
bulk of the stellar mass, formed over the course of a galaxy's history. 
Furthermore, \citet{Cortese:2012}, using UV and mid-IR star formation rates, 
concludes that optical colours are simply not able to distinguish between 
actively star-forming and passive galaxies.
 
The intermediate colours of GV galaxies has been interpreted as evidence for 
recent quenching of star formation \citep{Salim:2007}. It has been suggested
that, to explain the existence of 
such populations, galaxy transformations from blue 
to red must occur on short time-scales ($<1 \Gyr$, \citealt{Salim:2014}), but 
this picture is still controversial. 
\citet{Schawinski:2014} showed that the colour distributions of GV 
early-type galaxies can be modelled by a quenching 
time scale of $\sim 100 \Myr$, while for late-type galaxies a longer timescale,
$\sim 2.5 \Gyr$, is required. 
Recently, \citet{Crossett:2016} studied the impact of group environment 
on star formation in galaxies, finding that group galaxies have shorter 
quenching
time-scales ($<1\Gyr$) than non-grouped galaxies ($\sim2\Gyr$), thus 
demonstrating that environment plays an important role in the quenching process.
From the theoretical point of view, \citet{Trayford:2016} investigated the 
evolution and the origin of galaxy colours in the EAGLE cosmological 
hydrodynamical simulation \citep{Schaye:2015, Crain:2015}, and found a 
characteristic time scale for galaxies to cross the GV of $\sim 2 \Gyr$ which 
is independent of both galaxy mass and the physical mechanism responsible for
quenching. They found that galaxies with $M_{*}\leq10^{10}\Msol$ in the RS are
satellites that got their star formation suppressed by ram-pressure stripping in
the outskirts of more massive halos, while RS galaxies with 
$M_{*}\gtrsim 10^{10}\Msol$ are red, due to the feedback from their central 
supermassive black hole. 

Another scenario has been proposed out of the observations by
\citet{Salim:2012} and \citet{Fang:2012}: by studying ultraviolet morphology and star 
formation histories, these works propose that the GV may be static: most galaxies 
in the GV are not rapidly moving through it, rather, it is the green valley that moves 
slowly towards lower specific star forming rates. 

Moving from the BC to the RS is not the only way for a galaxy to be green; some
galaxies may reach the GV from the RS following the resumption of gas 
accretion from the intergalactic medium \citep{Thilker:2010}, or they may be
passive galaxies that recently underwent a minor, gas-rich merger 
\citep{Kaviraj:2009}.

Green valley galaxies are not a quiescent version of the star-forming galaxies,
because they are more centrally concentrated than galaxies in the 
\textit{main sequence}, which is a region in the stellar mass-specific star
formation rate (sSFR) diagram where most star-forming galaxies are found
\citep{Brinchmann:2004, Salim:2007,Peng:2010,Elbaz:2011,Leitner:2012}.
Therefore, the action of simple fading mechanisms, such as gas exhaustion 
or gas starvation, is not able to fully explain why galaxies could leave the 
main sequence \citep{Salim:2014}. 
Red galaxies are typically found in high galaxy density regions, 
so the environment plays a role in galaxy quenching. In addition, 
AGN feedback is a viable option 
\citep{Nandra:2007, Hasinger:2008, Silverman:2008, Cimatti:2013}. 
Recently, \citet{Schawinski:2014} concluded that green
early-type galaxies require a
scenario where both the gas supply and the gas reservoir are destroyed 
instantaneously, with rapid quenching accompanied by a morphological 
transformation from disc to spheroid. This gas reservoir destruction could be 
the consequence of a major merger which, alongside the burst in star formation,
fuels the accretion onto the central black hole, resulting in an upsurge of AGN 
feedback. In contrast, green late-type galaxies are consistent with
a scenario where the cosmic supply of gas is shut off, perhaps at a critical 
halo mass, followed by a slow exhaustion of the remaining gas over several 
$\Gyr$, driven by secular and/or environmental processes. 

Observational results suggest several models of quenching. Therefore, to 
investigate which quenching mechanisms are more important and the role of the 
environment, a comparison of GV galaxies in different environments can 
provide important clues to constrain the proposed scenarios. In this paper,
we perform a comparative analysis of a number of properties of galaxies in
the field, groups and clusters, searching for systematic differences 
in the GV, and also in the RS and BC, that could provide clues as to 
the environmental impact on the evolution of GV galaxies.  
This paper is organised as follows: in Sec. \ref{samples} we describe the 
samples of galaxies in X-ray clusters, groups and field; in Sec. 
\ref{results} we present our analyses; we discuss our results
in Sec. \ref{conclu}.

\begin{figure}
\includegraphics[width=8cm]{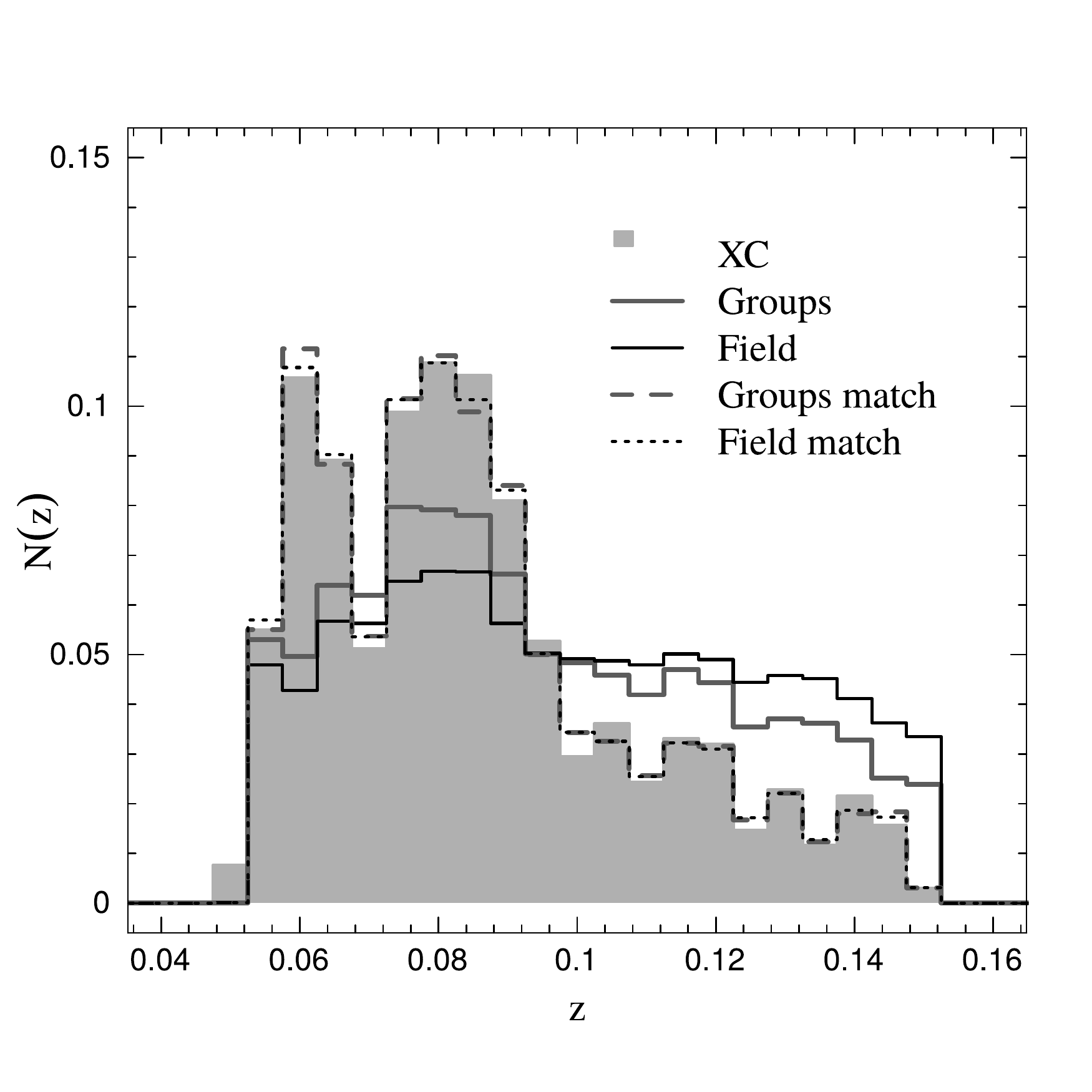}
\caption{Normalised redshift distribution of our samples of galaxies in 
X-ray clusters ({\em shaded grey area}), groups ({\em grey solid line}), and 
field ({\em black solid line}). {\em Dashed} and {\em dotted} lines show the 
redshift distribution of galaxies in groups and in the field, respectively, 
restricted to have redshift distributions similar to that of galaxies in X-ray 
clusters as described in the text. Kolmogorov-Smirnov tests comparing these 
distributions with that of the X-ray cluster galaxies confirm that they 
are consistent with being drawn from the same underlying distribution.}
\label{fig:z}
\end{figure}

\begin{figure}
\includegraphics[width=9cm]{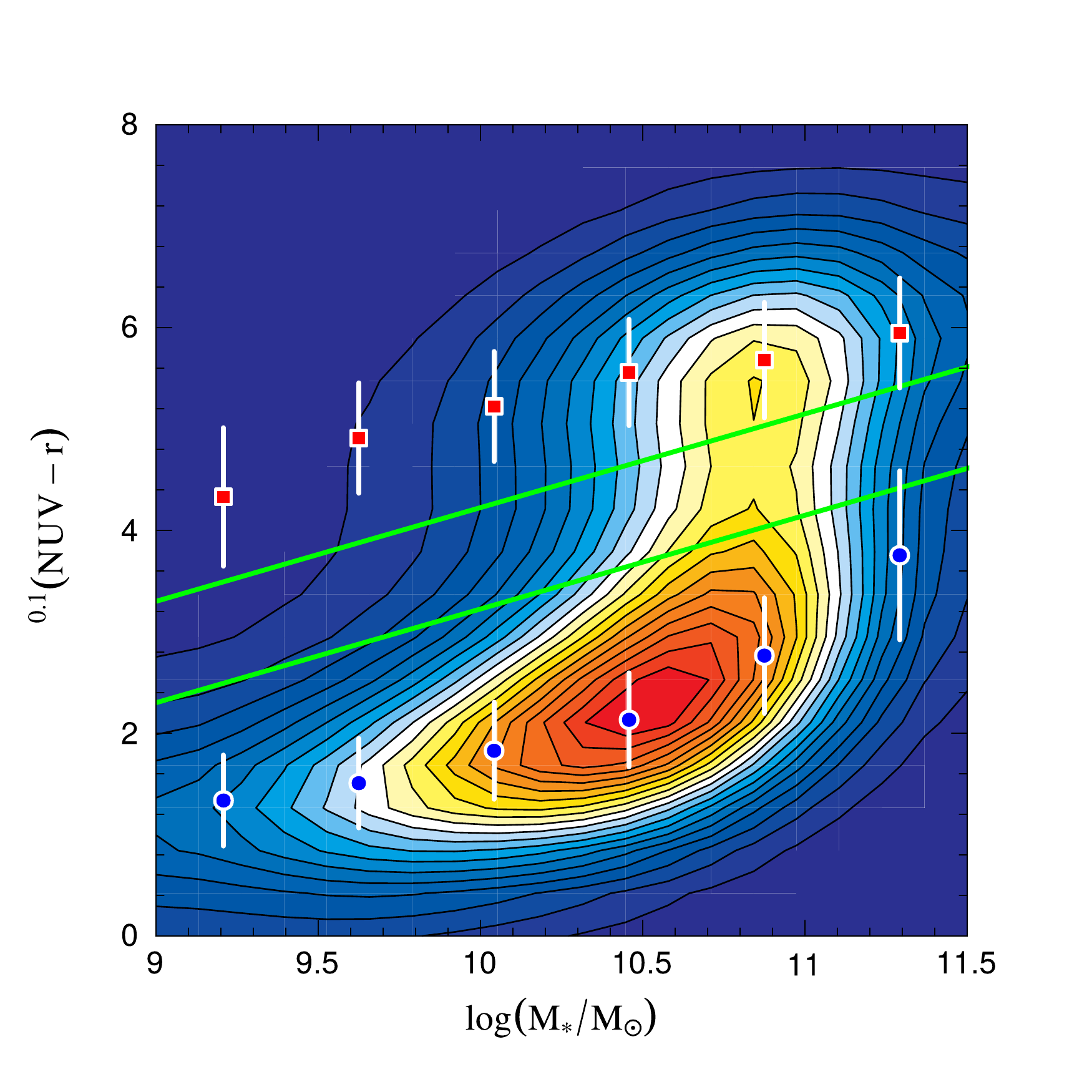}
\caption{$^{0.1}(NUV-r)$ colour-stellar mass diagram for MGS DR7 galaxies
restricted to $z\le0.15$. Blue circles and red squares are the centres of 
the Gaussian functions that fit best the blue and red populations, respectively.
Error bars are the corresponding widths. Green lines bound the region we 
consider as the green valley in this paper. See text for details.
}
\label{fig:cmassd}
\end{figure}

\section{The samples of galaxies}
\label{samples}

All samples of galaxies used throughout this paper are drawn from the 
Main Galaxy Sample (MGS, \citealt{Strauss:2002}) of the Sloan Digital Sky 
Survey's (SDSS, \citealt{York:2000}) Seventh Data Release (DR7, \citealt{dr7}). 
Since the GV is better defined in terms of UV-optical 
colours, we restrict our analysis to SDSS galaxies that have $NUV$ magnitude measured by the Galaxy Evolution Explorer (GALEX).
We take UV data from the final data release of GALEX, GR6/7
\footnote{http://galex.stsci.edu/GR6}. For each SDSS galaxy, we search for the
nearest GALEX source that is within $4\arcsec$. To avoid the edge
of the detectors, where GALEX photometry and astrometry degrade, we further 
restrict to matches that are within $0.6^{\circ}$ of the GALEX field of view 
centres. In all cases, we use the $NUV$ magnitude from the deepest available
imaging survey of GALEX (AIS, MIS and DIS). 
SDSS's $r-$band magnitudes have been corrected for Galactic
extinction using the maps by \citet{sch98}. $NUV$ magnitudes have been 
corrected for Galactic extinction using the same extinction maps and assuming
$A_{NUV}=8.87E(B-V)$ \citep{Chilingarian:2012}. We compute absolute magnitudes 
assuming a flat cosmological model with parameters $\Omega_0=0.3$, 
$\Omega_{\Lambda}=0.7$ and a Hubble's constant $H_0=100~h~\kms~{\rm Mpc}^{-1}$. 
We have $K-$corrected $NUV$ and $r-$band magnitudes to a redshift $z=0.1$, which
roughly corresponds to the peak of the redshift distribution of MGS galaxies,
by means of the ZEBRA\footnote{Zurich Extragalactic Bayesian Redshift 
Analyzer, {\tt http://www.exp-astro.phys.ethz.ch/ZEBRA}} code 
\citep{Feldmann:2006}. 
We refer to these band-shifted magnitudes with the superscript $0.1$.
UV minus optical colours are affected by dust. Following 
\citet{Wyder:2007}, we correct the $^{0.1}(NUV-r)$ colour using the empirical 
dust-SFH-colour relation derived by \citet{Johnson:2006}. Firstly, we compute
the $FUV$ attenuation $A_{FUV}$ as a function of $D_n(4000)$ 
and the uncorrected $^{0.1}(NUV-r)$ from the fits in Table 1 of \citet{Johnson:2006}. 
Secondly, we calculate the $NUV$ and $r$ assuming $A_{NUV}=0.81A_{FUV}$
and $A_r=0.35A_{FUV}$, which are derived from \citet{Calzetti:2000}.
Unless otherwise specified, colours used throughout this paper are 
dust-corrected. Stellar masses $Dn(4000)$ and specific star formation rates 
(sSFR) have been taken from the MPA-JHU catalogue 
\citep{Kauffmann:2003,Brinchmann:2004}. All magnitudes are in the AB system.

\begin{figure*}
\begin{minipage}[c][12.5cm][t]{.33\textwidth}
  \centering
  \includegraphics[width=6cm,height=6cm]{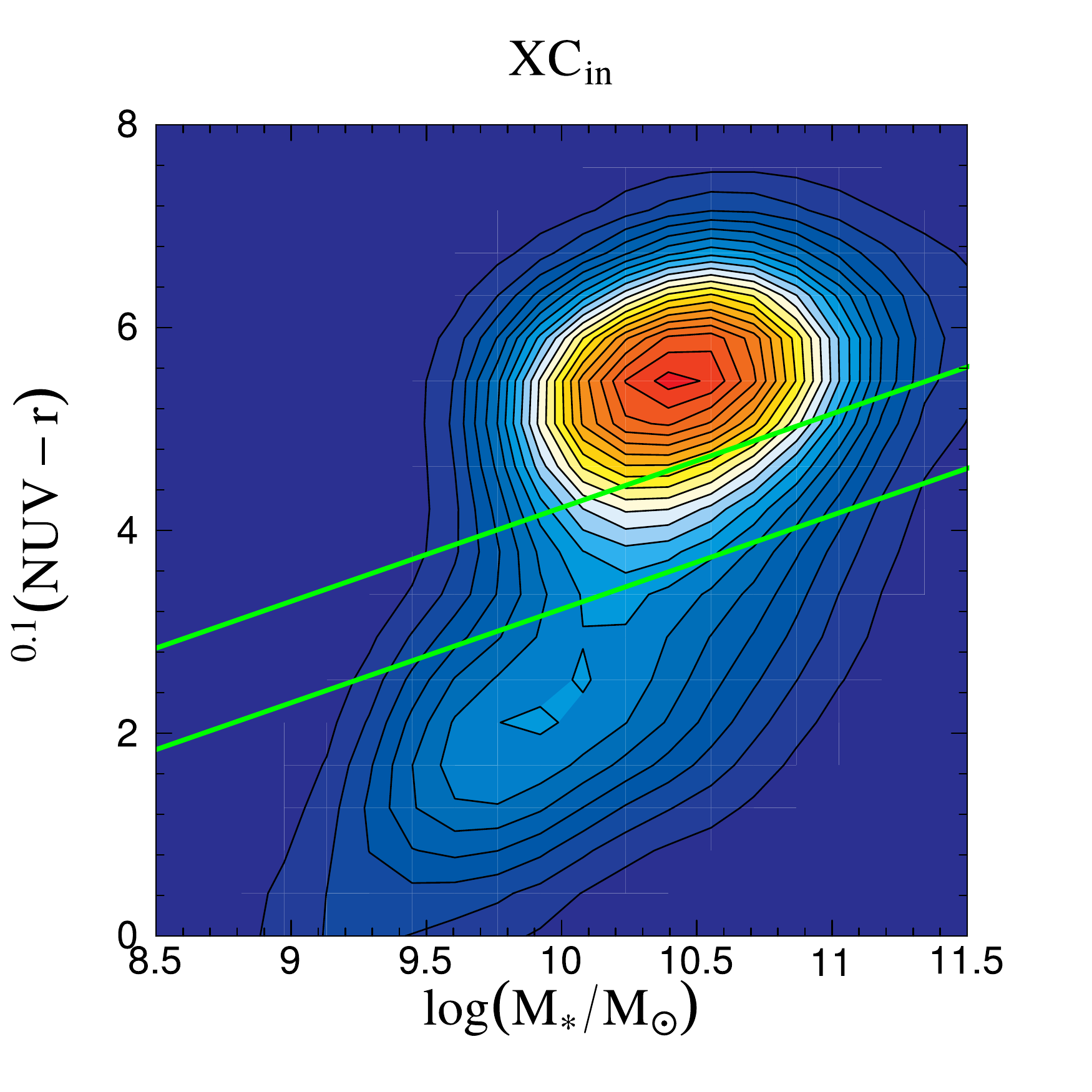}
  \includegraphics[width=6cm,height=6cm]{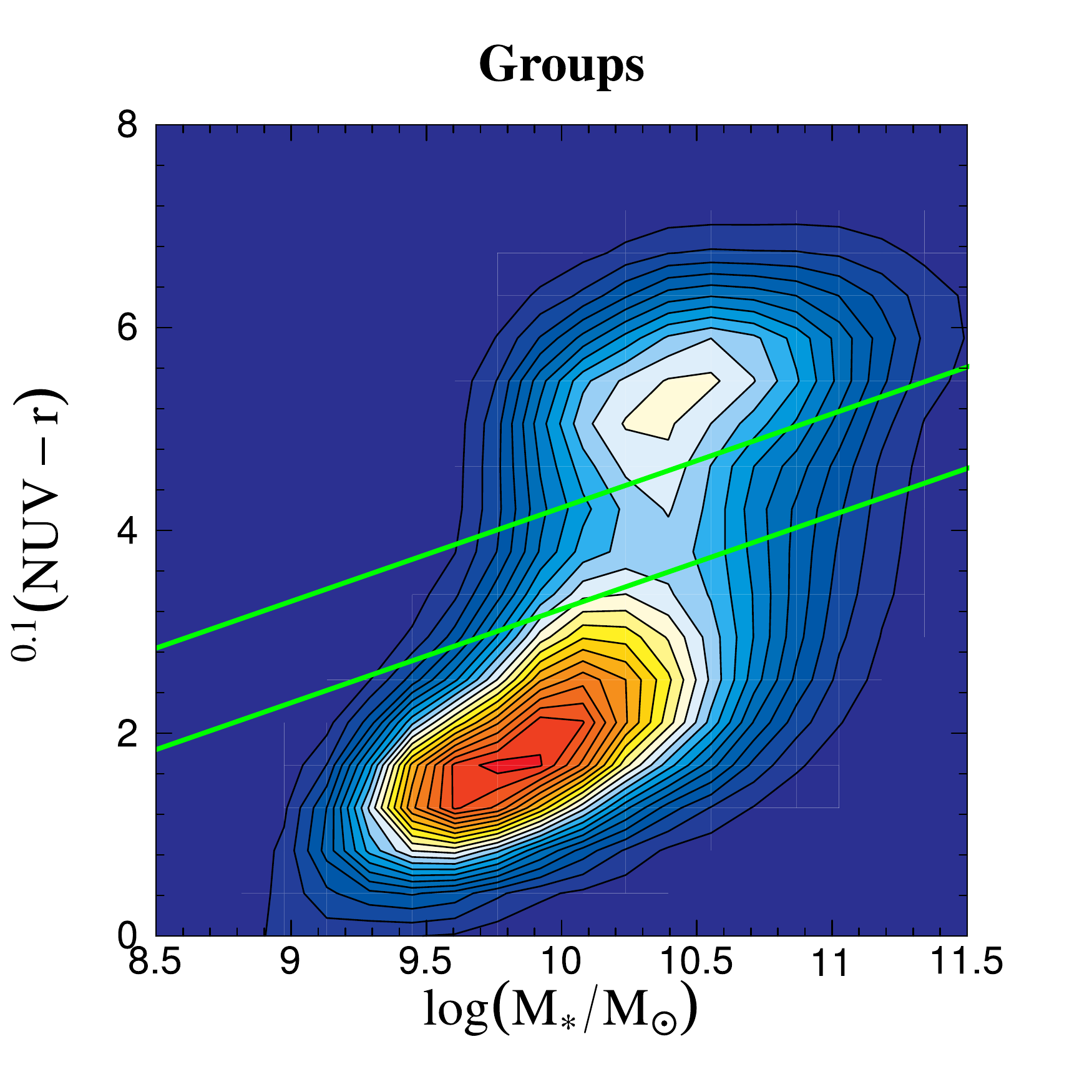}
\end{minipage}
\begin{minipage}[c][12.5cm][t]{.33\textwidth}
  \centering
  \includegraphics[width=6cm,height=6cm]{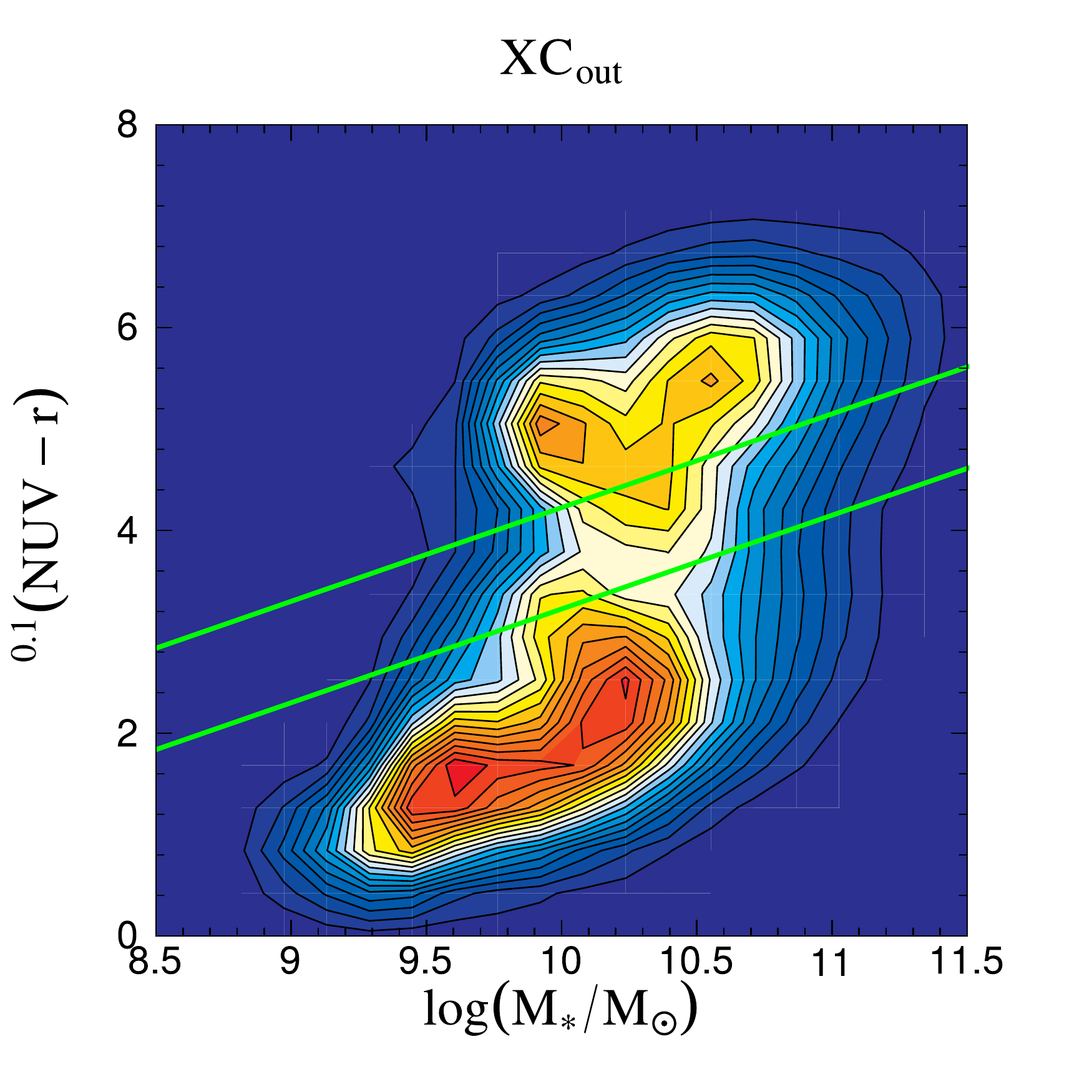}
  \includegraphics[width=6cm,height=6cm]{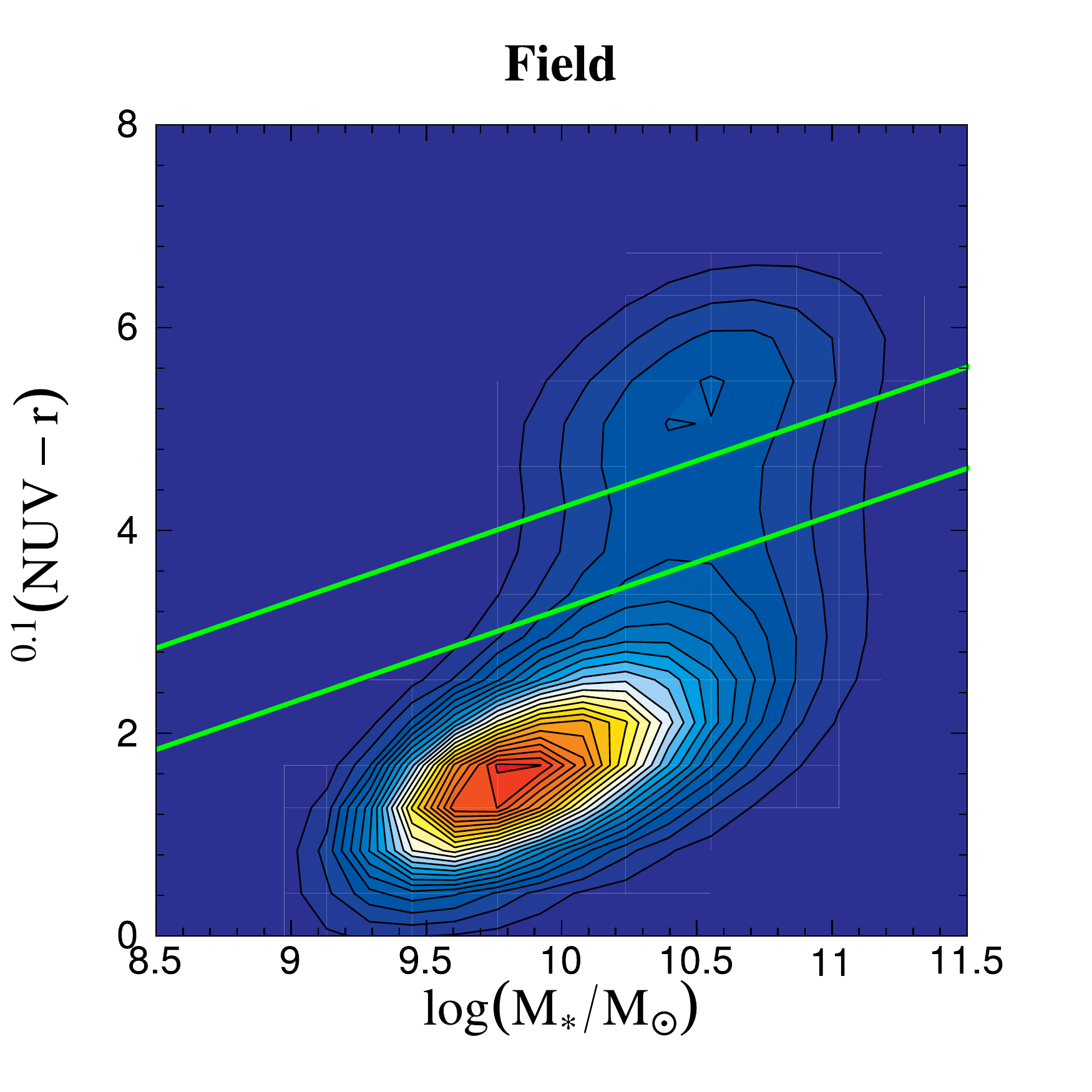}
\end{minipage}
\begin{minipage}[c][12.5cm][t]{.3\textwidth}
  \vspace{0.2cm}
  \centering
  \includegraphics[width=4.5cm,height=12cm]{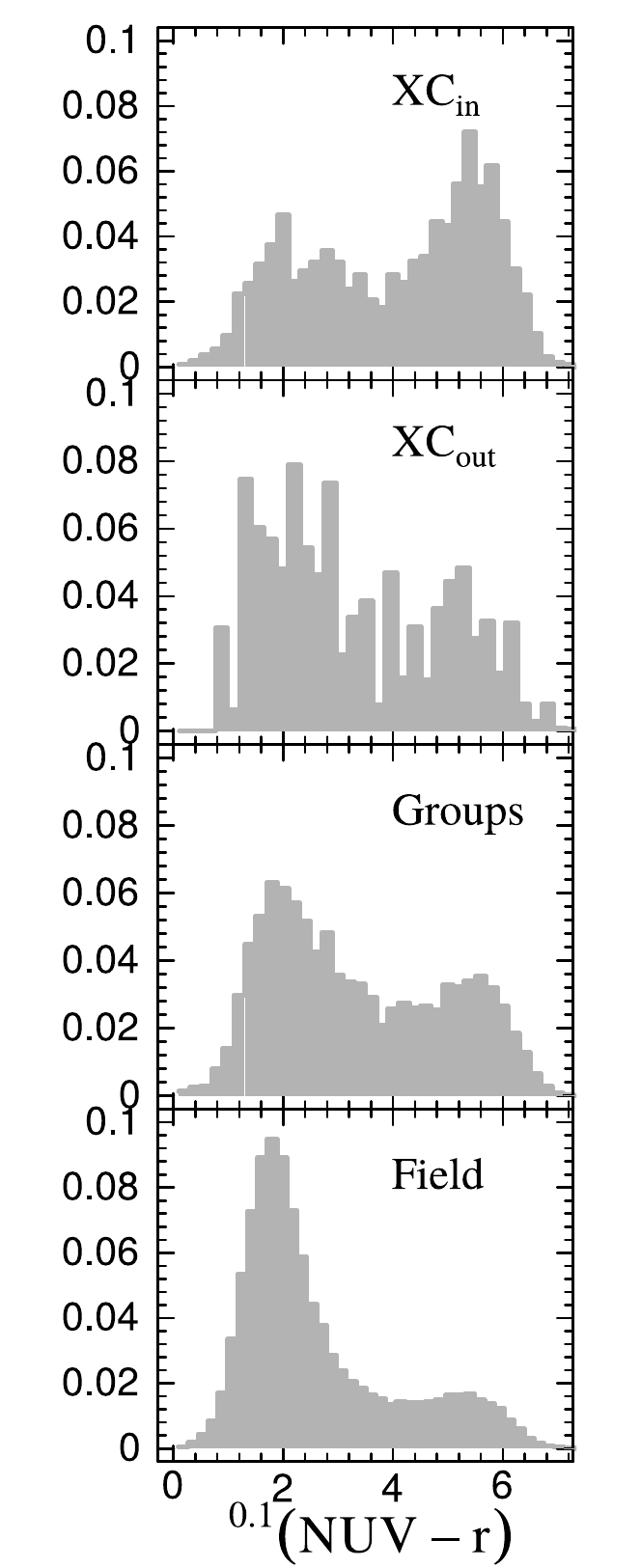}
\end{minipage}
\caption{UV-optical colour-mass diagram. Contours indicate
increasing density from {\em blue} to {\em red}. {\em Top left:} galaxies in
the inner regions of X-ray clusters. {\rm Top right:} galaxies in the outskirts
of X-ray clusters. {\em Bottom left:} galaxies in groups. {\em Bottom right:}
field galaxies. Horizontal {dashed lines} define the three sequences (see text).
To the right side of each panel the corresponding $^{0.1}(NUV-r)$ distributions
are shown.}
\label{fig:cmd2}
\end{figure*}

The green valley is the region that lies between the red sequence and the 
blue cloud in the colour-magnitude or the colour-stellar mass diagram. 
These two sequences bounding the green valley depend on mass (or luminosity),
and so does the GV. We show our definition of green valley
in the colour-mass diagram of Fig.\ref{fig:cmassd}, where we include all MGS 
DR7 galaxies restricted to $z\le0.15$ which have measured $NUV-$band photometry.
To define the GV in this diagram,
we proceed as follows: 1) We split the galaxies into seven bins of stellar mass,
ranging from $10^9M_{\odot}$ to $3.2\times10^{11}M_{\odot}$; 2). For each 
stellar mass bin, we consider the $^{0.1}(NUV-r)$ colour distribution, and
fit Gaussian functions separately to both the blue and the red populations.
To avoid contamination from the GV in the fits, we use only 
those points in the colour distribution that lie no further red (blue) than
0.5 mag of the blue (red) peak. We show in Fig.\ref{fig:cmassd} as blue/red 
points the centres of the Gaussian functions and as white error-bars the
corresponding widths; 3) We fit straight lines to the position of the peaks of 
the blue and red populations as a function of mass. It is clear from 
Fig.\ref{fig:cmassd} that the blue peaks could be described by a different
functional form, but we consider this to be beyond the scope of this paper;
4) We define the GV on the colour-mass diagram as those galaxies lying in 
the region defined by:
\begin{equation}
^{0.1}(NUV-r)(M_{*})=0.92\times M_{*}-5.52\pm0.5.
\end{equation}
The slope of these straight lines is the mean between the slopes
of the fits to the blue and red peaks. We have chosen the zero point in
order to define a 1 mag wide strip bounded by the two sequences.
This also defines the passive sequence (PS) and the star-forming sequence 
(SFS), as those galaxies lying above or below the GV, respectively.

Our goal is to perform a suitable comparison of GV galaxies in 
different environments. Any differences in the populations of GV galaxies due
to different environmental effects could be spotted only if the samples
of GV galaxies in the environments explored here can be directly compared.
Our samples of galaxies in X-ray clusters, groups, and in the field were 
carefully selected to have similar redshift distributions.  

\subsection{The sample of galaxies in X-ray clusters}
Our sample of galaxies in X-ray clusters has been drawn from two sources:
the C-P04-I sample of \citet{Coenda:2009}, and the C-B00-I sample of 
\citet{Muriel:2014}. The former was constructed from the ROSAT-SDSS Galaxy 
Cluster Survey of \citet{Popesso:2004}, and the latter from the
the Northern ROSAT All-Sky Galaxy Cluster Survey of \citet{Noras:2000},
and comprise 49 and 55 clusters, respectively, in the redshift range
$0.05\le z\le 0.14$. Galaxies belonging to the clusters in these samples
were identified using the same procedure over the DR7 MGS. The authors
identified cluster members in two steps. First, they used a friends-of-friends 
(\citealt{H&G:1982}, hereafter {\em fof}) algorithm that uses the linking 
parameters and modifications introduced by \citet{Diaz:2005}. Secondly, they 
performed eyeball examinations of the structures detected by \textit{fof}. 
From the redshift distribution of galaxies, they determined the galaxy
members from line-of-sight extension of each cluster.
By visually inspecting every cluster, the authors excluded systems that 
have two or more close substructures of similar size in the plane of the 
sky and/or in the redshift distribution. Using the galaxy members identified
this way, the authors computed the cluster physical properties: 
line-of-sight velocity dispersion, virial radius and mass. 
Clusters in the C-P04-I sample have mean virial mass of 
$7\times10^{14}~M_{\odot}$, mean virial radius of $1.75\mpc$ and
mean line-of-sight velocity dispersion of $715\kms$.
Clusters in the C-B00-I sample have mean virial mass of 
$9\times10^{14}~M_{\odot}$, mean virial radius of $1.83\mpc$ and
mean line-of-sight velocity dispersion of $820\kms$.

\begin{figure}
\includegraphics[width=9cm]{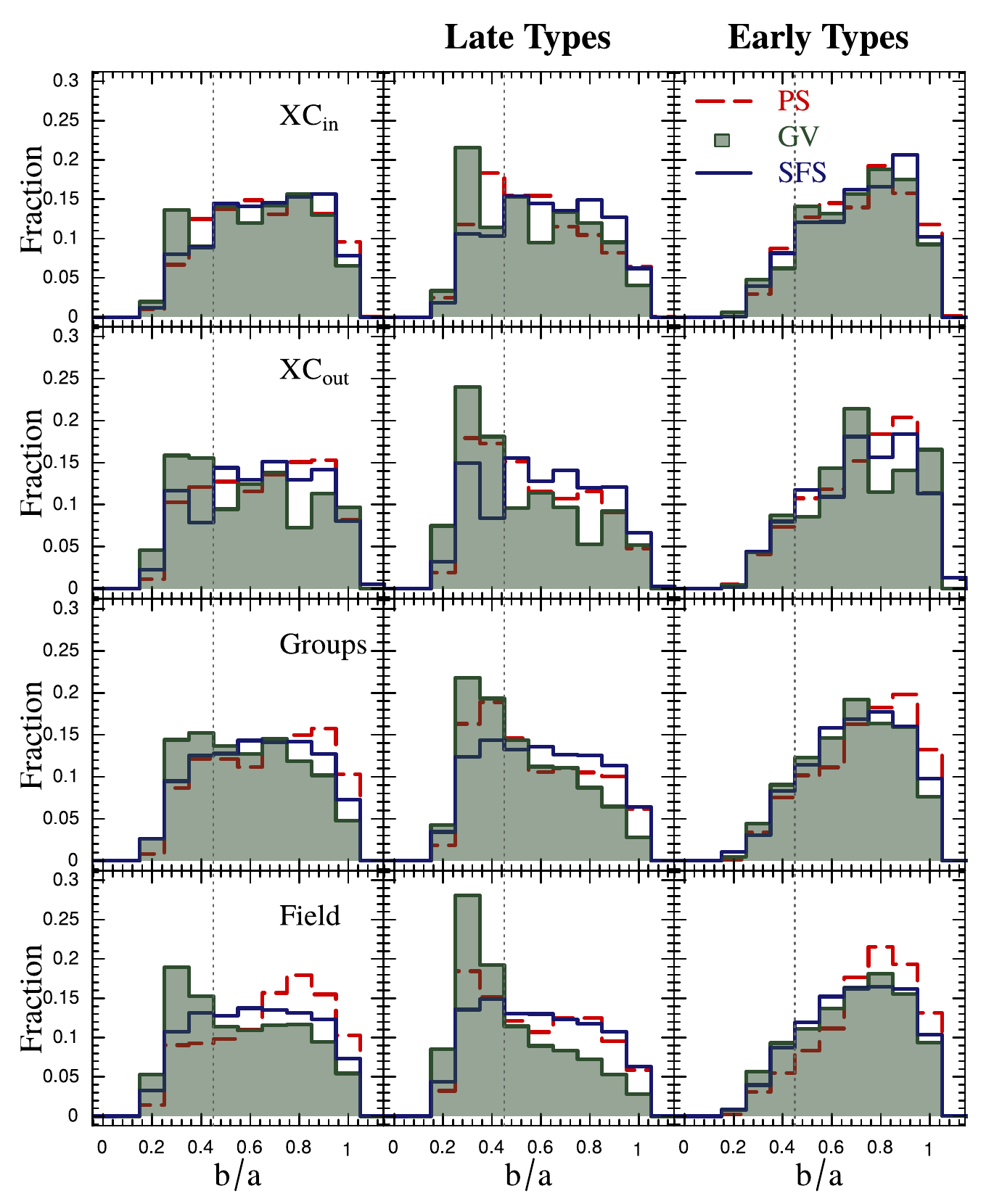}
\caption{Normalised distribution of $r$-band minor-to-major axis ratio $b/a$ 
as a function of the environment. Dashed lines show the cut $b/a\geq0.45$ we 
use to remove dusty edge-on star-forming galaxies. Galaxies in the PS
are shown in {\em red lines}, the GV as a {\em shaded green area}, and the SFS 
in {\em blue lines}.}
 \label{fig:ba}
\end{figure}

We include in our sample of galaxies in X-ray clusters, all galaxies
in the C-P04-I and C-B00-I clusters that have GALEX $NUV$ photometry
with the restrictions explained above.
Among the three environments explored in this paper, the sample
of galaxies in X-ray clusters is, by far, the sparsest, accounting for 4040
galaxies. Thus, 
this sample will, henceforth, define the redshift distribution that 
our samples of field and group galaxies must have. The redshift distribution 
of cluster galaxies can be seen in Fig. \ref{fig:z}.

\subsection{The sample of galaxies in groups}
We draw our sample of galaxies in groups from the sample of groups 
identified by Zandivarez \& Mart\'inez (2011, hereafter ZM11) \nocite{ZM11} in the SDSS 
DR7 MGS. Briefly, they use a standard {\em fof} algorithm  
to link galaxies into groups by means of a redshift-dependent linking length. 
ZM11 split merged systems and cleaned up spurious member detection through
a complementary identification procedure, using a higher 
density contrast in groups with at least 10 observed members \citep{Diaz:2005}.
Virial masses are computed from the virial radius of the systems and the 
velocity dispersion of member galaxies \citep{Limber:1960,Beers:1990}. 
Groups in the ZM11 sample have mean virial mass 
$2.1\times10^{13}M_{\odot}$, mean virial radius $0.9\mpc$ and mean
line of sight velocity dispersion of $193\kms$.
The catalogue of ZM11 includes 15961 groups with more than four members, 
adding up to 103342 galaxies. 
We refer the reader to ZM11 and references 
therein for further details of group identification. 

From the original sample of ZM11, we randomly choose among galaxies with 
$NUV$ photometry, with a Monte-Carlo algorithm that results in a subsample of 
group galaxies that has a similar redshift distribution of that of our sample 
of galaxies in X-ray clusters. This subsample comprises 17797 galaxies.
The redshift distribution of this sample of group 
galaxies is also shown in Fig. \ref{fig:z}.

\subsection{The sample of field galaxies}
Out of the remaining MGS galaxies that were not included as cluster galaxies 
or as group galaxies by ZM11, and that have NUV photometry, we 
randomly select as many objects as we can, as long as they have similar 
redshift distribution to the sample of galaxies in X-ray clusters, using the 
same algorithm we use for constructing the sample of group galaxies.
This subsample comprises 92246 galaxies.
The redshift distribution of this sample of field galaxies is compared to that
of X-ray cluster galaxies and group galaxies in Fig. \ref{fig:z}. 
The samples of clusters and groups used in this paper, 
are, by construction, not volume complete. Therefore, there are galaxies that
belong to clusters and groups that were not identified as such, and thus,
they might be contaminating our sample of field galaxies. There is no safe way
to exclude these interlopers from our field sample; however, they should be 
outnumbered by actual field galaxies. Our results regarding field galaxies 
should be read taking this into account.


\section{Comparing PS, GV and SFS galaxies in different environments} 
\label{results}

This paper aims to perform a comparison of the properties of galaxies in the PS,
the GV and the SFS, in four discrete environments: field, groups, 
and the two in X-ray clusters. The environment itself varies dramatically inside X-ray 
clusters; the outskirts of these systems are very different from their innermost
regions. Therefore, in our analyses, we further distinguish between galaxies 
located in two regions within clusters: cluster core ($r/r_{200}\leq0.5$, 
hereafter XC$_{\rm in}$), and the outskirts of clusters ($r/r_{200}>0.5$, hereafter
XC$_{\rm out}$), where $r_{200}$ is the radius that encloses a density 200 times 
the mean density of the Universe.

In Fig. \ref{fig:cmd2} we show the UV-optical colour-stellar mass diagram 
of galaxies in the four environments, and the normalised distributions of 
$^{0.1}(NUV-r)$ colour. We include in this figure only those
galaxies that have minor to major axis ratio $b/a>0.45$. The justification 
for this cut-off is explained later in the section.
In this figure, a gradual transition is observed, from the
field, which is largely dominated by SFS galaxies, to the inner regions
of clusters, where most of the galaxies are in the PS.

We use the morphological classifications taken from the Galaxy Zoo Project 
(\citealt{Lintott:2008}) to analyse early- and late-type galaxies.
Briefly, each galaxy in the Galaxy Zoo Project received several, independent 
morphological classifications, each performed by a different user. These
classifications were processed into raw likelihoods $P$ for every galaxy
of a particular morphological type (elliptical, spiral, merger, and 
{\em don't know}), directly from the ratio of the number of classifications as 
being of that type to the total. Throughout this paper we use the probabilities
$P_E$ (elliptical) and $P_{SP}$ (spiral) corrected after the de-biasing procedure of 
\citet{Lintott:2011}. The elliptical class contains galaxies with elliptical 
morphology and the majority of S0 galaxies, therefore we refer to it henceforth
as the early-type class (ET). The spiral classification contains galaxies with 
different directions and orientation of the spiral arms, and we refer to it as 
late-type galaxies (LT). As \citet{Bamford:2009}, we use the raw measures 
likelihood to weight each galaxy in the statistics by its probability 
of being either elliptical or spiral, instead of classifying all galaxies
into two types defined by a certain likelihood threshold. Hereafter, when we 
mention our results concerning early or late types, we mean that in the
statistics we have used the likelihood $P_E$, or $P_{SP}$, as a weighting.
This has the advantage that it retains more information than 
thresholding, and all galaxies can be included, but it cannot provide 
classifications for individual objects.

We show in Fig. \ref{fig:ba} the normalised distribution of the best model 
$r$-band ratio of the axis $b/a$, as provided by the SDSS database, as a 
function of the environment, for all late- and early-type galaxies. Edge-on 
galaxies have $b/a=0$ which means an inclination of $90^{\circ}$.
Late-type green galaxies are predominantly edge-on galaxies, independently of 
the environment considered. Edge-on systems darken the ultraviolet radiation 
due to the presence of dust, shrouding the star formation and, therefore, 
affecting the colour of the galaxy itself. 
Although we have corrected colour for dust, hereafter we only consider 
galaxies with $b/a\geq0.45$, to avoid objects that could be strongly affected
by dust and thus have unreliable UV photometry.
This threshold is shown in dashed line in Fig. \ref{fig:ba}. 
Other authors have taken into account this source of bias. For instance,
\citet{Fang:2012} use a threshold of $b/a=0.65$ to select face-on galaxies.
Imposing a cut-off in axis ratio implies that more spirals are removed 
than ellipticals. Despite Fig. \ref{fig:ba} shows that the distribution of axis 
ratios has a small variation with environment and the star formation classification,
we can safely assume this cut will not introduce major biases in our comparative 
analyses below.

\subsection{Morphology}
\begin{figure}
\includegraphics[width=8cm]{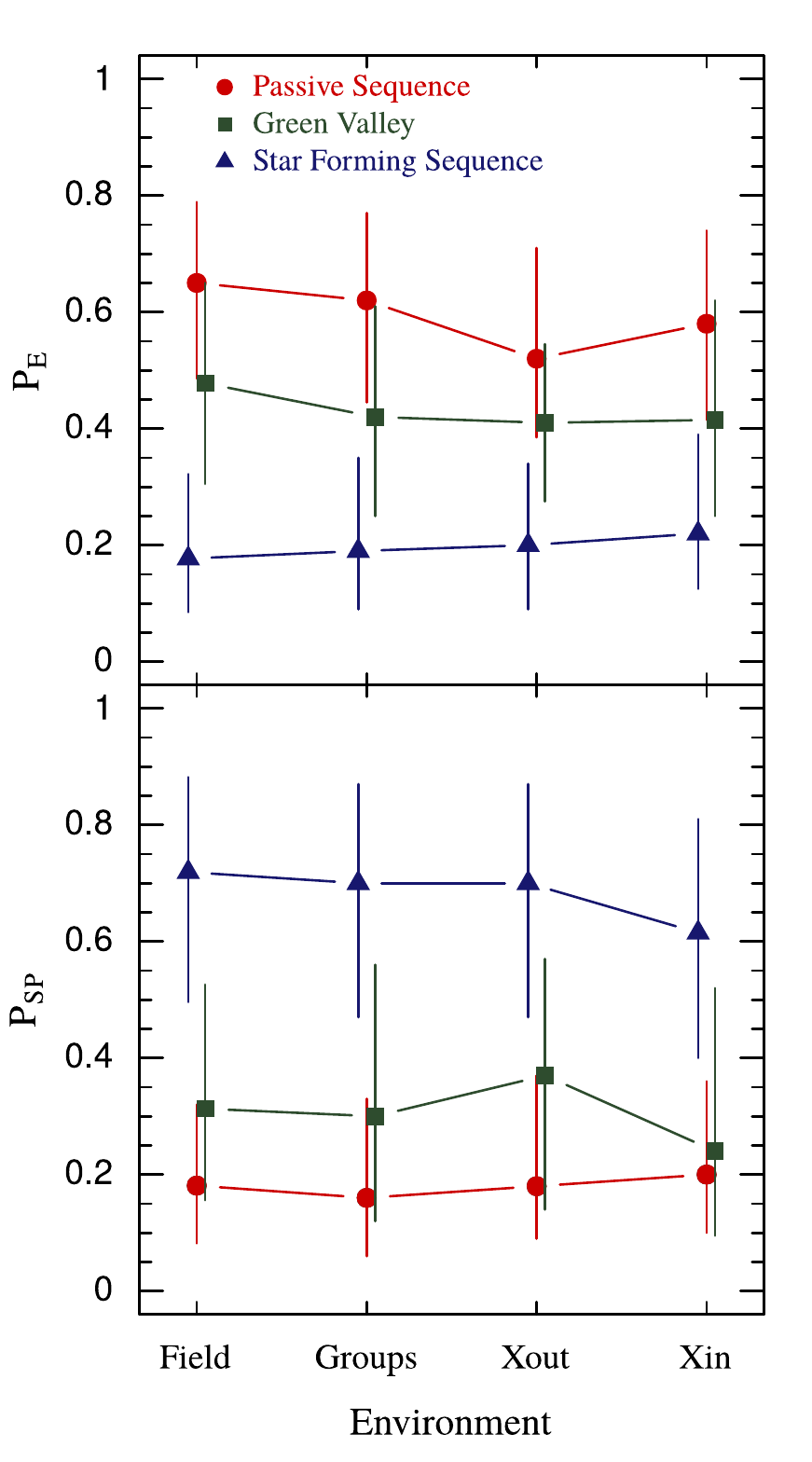}
\caption{
Weighted median probabilities of being Elliptical (upper panel) and
Spiral (lower panel) as a function of environment. Vertical error bars
are the 25\% and 75\% percentiles. GV (SFS) points have been shifted in 0.05
(-0.05) on the x-axis.
}
 \label{fig:zoo}
\end{figure}

The upper (lower) panel of Fig. \ref{fig:zoo} shows the weighted median probabilities
$P_E$ ($P_{SP}$) of galaxies in the different environments and sequences considered. 
Vertical bars represent the 25\% and 75\% percentiles. We include in this figure only  
galaxies more massive than $10^{9.8} M_{\odot}$.  
The justification for this cut-off is explained in the next section.
Galaxies in the PS have the highest (lowest) median values of $P_E$ ($P_{SP}$), and 
they are almost independent of the environment. GV galaxies have median values 
of $P_E$ and $P_{SP}$ in between those of galaxies in the SFS and the PS, although 
closer to the values of the PS. The morphology of GV galaxies is also nearly 
independent
of the environment, with the possible exception of the inner region of 
clusters, where we find that the median $P_{SP}$ in the GV and the PS are very 
similar. The intermediate morphologies of GV 
galaxies have been reported by other authors (e.g. \citealt{Schawinski:2014}). 
\citet{Schiminovich:2007} found that GV galaxies have S\'ersic indices that are 
halfway between those of galaxies in the PS and the SFS. \citet{Mendez:2011} 
also found that GV galaxies are intermediate between red and blue galaxies, in 
terms of concentration, asymmetry and morphological type. Our results confirm 
these findings, with the exception of galaxies in the inner region of clusters, 
where few green spirals are found. If there is an effect, it
might be an indication that, in the inner region of clusters, both the quenching 
and the morphological transformation times are shorter than in other environments. 

\subsection{Stellar mass}

\begin{figure*}
\includegraphics[width=15cm]{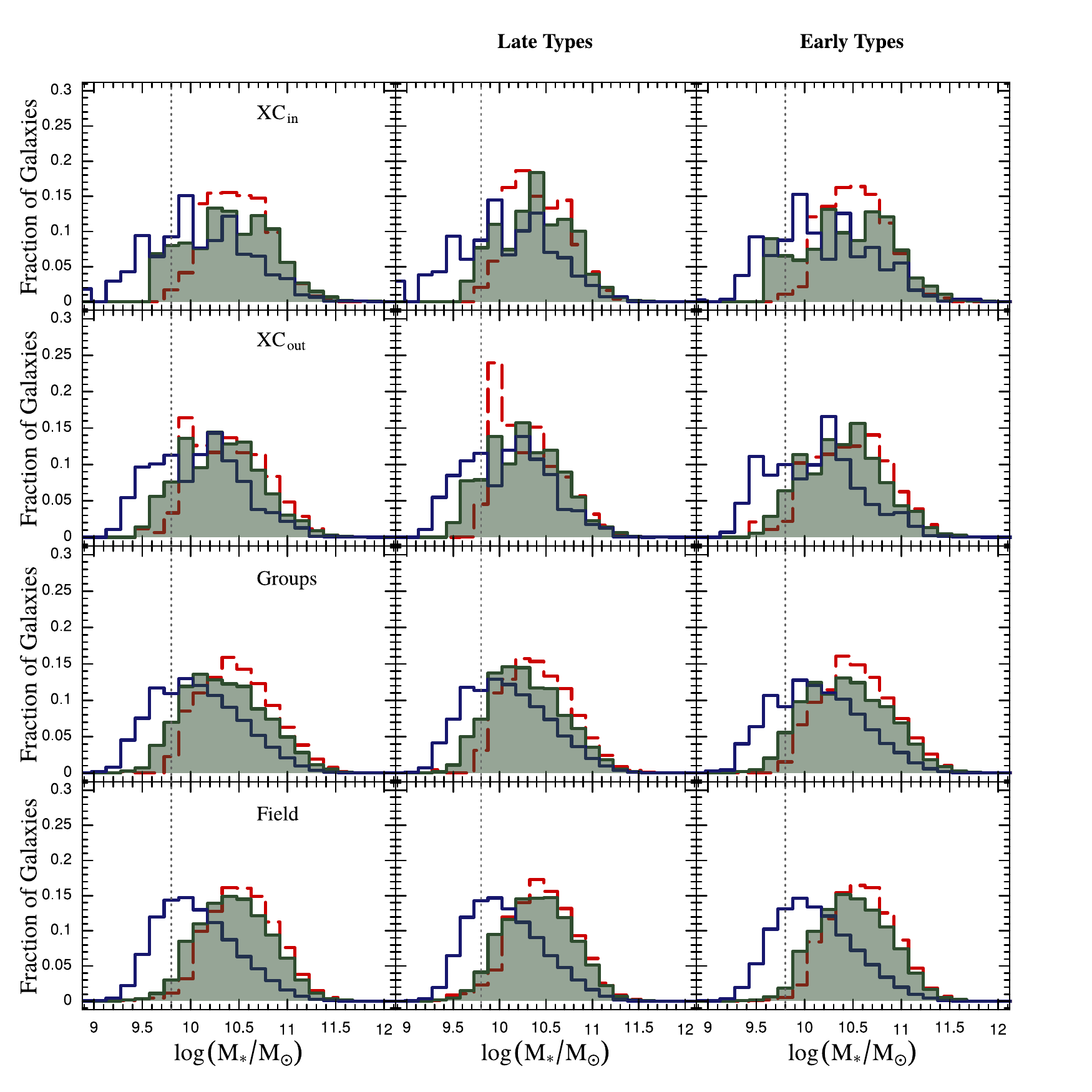}
\caption{Normalised distribution of stellar mass (MPA-JHU catalogue), for PS, GV 
and SFS. Colours are as in Fig. \ref{fig:ba}. {\em Left column:} all galaxies, 
{\em central column:} late-type galaxies, {\em right column:} early-type 
galaxies. }
\label{fig:masa}
\end{figure*}

In Fig. \ref{fig:masa} we compare the normalised stellar mass
distributions of galaxies in the PS, the GV and the SFS. 
First of all, we note that low stellar mass galaxies are only 
present in the SFS sub-sample. This is basically due to sample construction:
we require that all galaxies in our samples have $NUV$ photometry, which implies
that for a galaxy to be included in our samples, it must be brighter in $NUV$
as its stellar mass is smaller, so as mass decreases, the fraction of SFS
galaxies increases to the detriment of the other sequences. 
Although our samples are not volume-complete, the selection is identical 
across environments. Therefore, at fixed stellar mass, it is appropriate 
to compare the fraction of SFS, GV and PS in different environments.
Since there is a lack of galaxies at the low end of the distribution
in the GV and the PS, for subsequent analyses we only consider galaxies 
with $\log(M_{\ast}/M_{\odot}) \geq 9.8$, in order to have 
a fair comparison between environments at the low mass end. After this cut-off,
in every binned statistics we have more than 10 galaxies per bin. 
This threshold is shown in Fig. \ref{fig:masa} as vertical dotted lines. 
After the cut-offs in $b/a$ and stellar mass, the samples of field,
group and cluster galaxies, comprise 60553, 12943, and 3007 objects, respectively.

\begin{figure*}
\includegraphics[width=15cm]{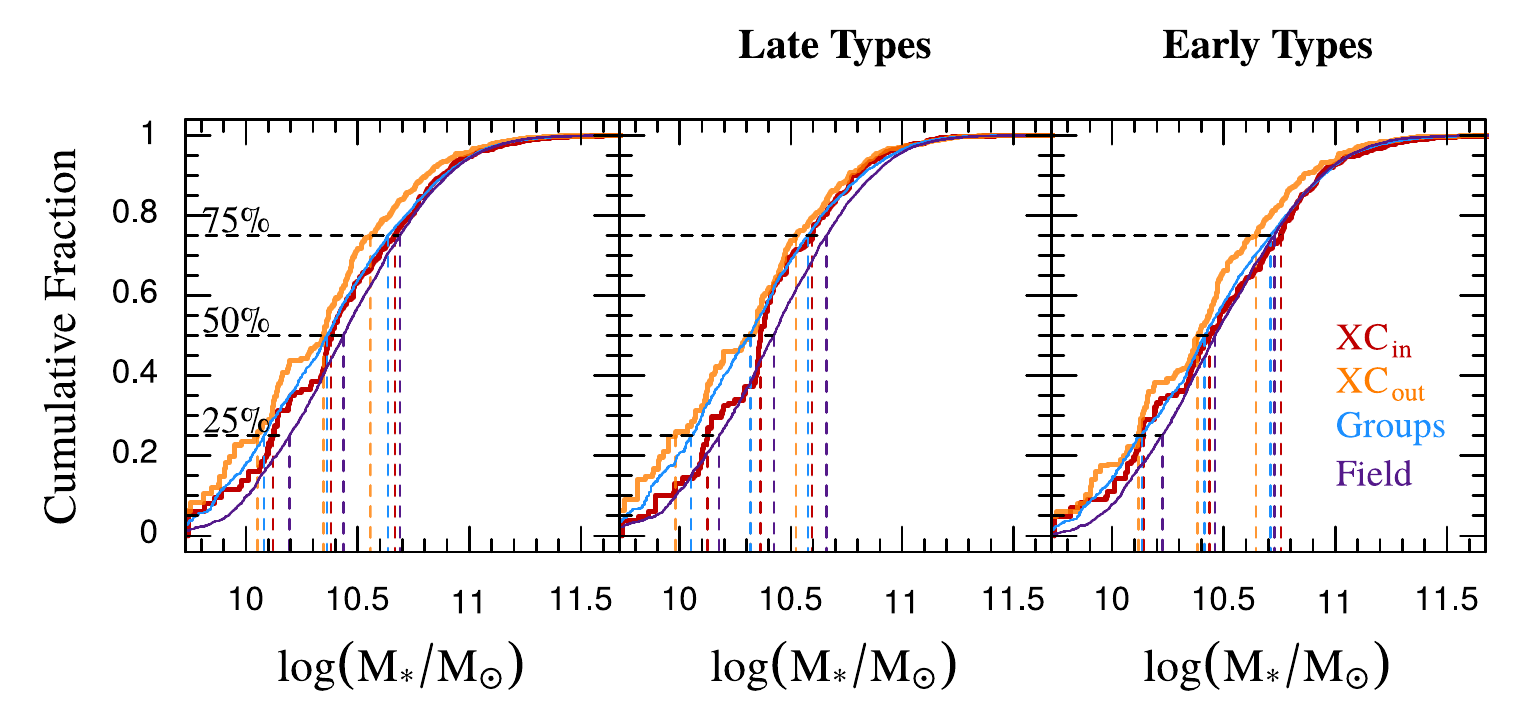}
\caption{Cumulative fraction of stellar mass for GV galaxies:
{\em left panel} corresponds to all galaxies, {\em central panel}
to late-types, and {\em right panel} to early-types. 
{\em Dashed lines} represent $25\%$, median and $75\%$ 
quartiles of the stellar mass.}
\label{fig:masaG}
\end{figure*}

We find that, with the exception of ET-GV galaxies 
in Groups and XC$_{\rm in}$ (60\% probability for the null hypothesis) there is no
sequence showing similar mass distributions in different environments. 
To further explore differences in the GV, in Fig. \ref{fig:masaG} we compare the 
stellar mass cumulative fraction of GV 
galaxies for the four environments considered. We note that, in
the field, GV galaxies are slightly more massive than in other environments. 
This difference is larger for LT galaxies; nevertheless, it is also observed 
for ET. If internal quenching is the main driver of the transformations of 
galaxies in the field, it will preferentially affect high mass galaxies 
\citep{Peng:2010}, while 
in groups or clusters the environmental effects will also transform lesser
massive galaxies. The highest cumulative fraction of low mass GV galaxies 
occurs in the outskirts of clusters ({\em orange line} in Fig. 
\ref{fig:masaG}), suggesting that this environment is particularly efficient
in transforming these galaxies. 

\subsection{The relative abundance of the sequences}

\begin{figure*}
\includegraphics[width=15cm]{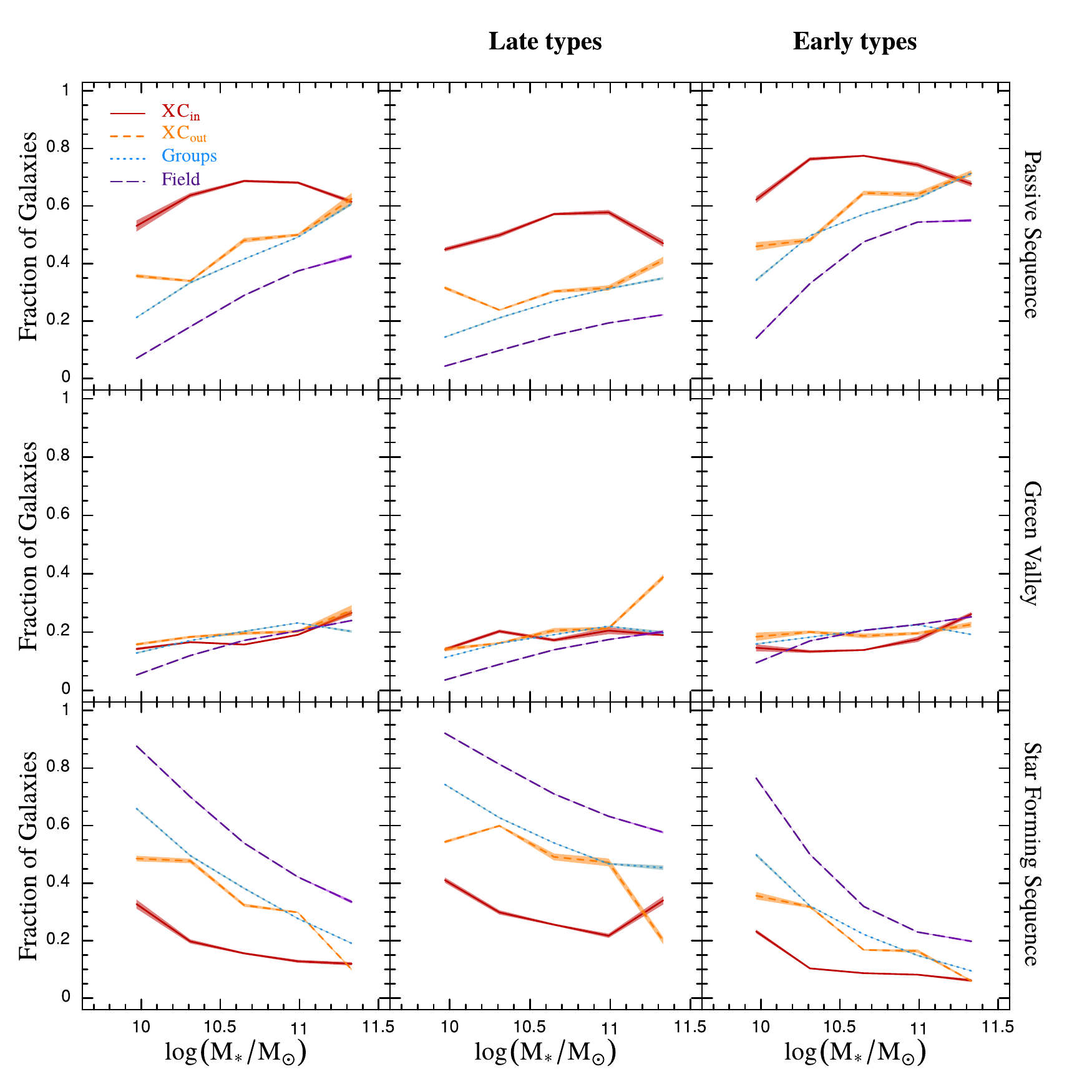}
\caption{Fraction of galaxies as a function of stellar mass and environment. 
{\em Left column:} 
all galaxies irrespective of their type, {\em central column:} late-type 
galaxies, and {\em right column} early-type galaxies. {\em Upper panels}
correspond to PS galaxies, {\em middle panels} to GV, while the 
{\em bottom panels} represent the SFS. 
Vertical error bars were computed by using the bootstrap resampling technique.}
\label{fig:frac}
\end{figure*}

The ability of the different environments to transform galaxies can be studied 
by analysing the fraction of objects with different levels of star formation
activity as a function of the stellar mass. In Fig. \ref{fig:frac}, 
we show the fraction of PS, GV, and SFS galaxies as a function of stellar mass 
and the environment. 
At fixed mass and environment, the sum of the fraction of galaxies 
in all three sequences equals 1. Left column shows the fraction of galaxies 
irrespective
of their morphological type, central column considers late types, and
right column early types. Recall that the number of galaxies remains the same
across columns, since late and early types fractions refer to the same 
galaxies weighted by their likelihood of being either late, or early types.
As an example of the robustness of the statistics in this figure, 
the bin with the lowest number of galaxies is the
highest mass bin of the XC$_{\rm out}$ environment, which contains 106 galaxies:
44 in the PS, 41 in the GV, and 21 in the SFS.
Hereafter, all errorbars are computed by 
the bootstrap resampling technique, unless otherwise specified. 
As expected, the fraction of galaxies in the PS grows with $M_{\ast}$ and the 
opposite behaviour is observed for galaxies in the SFS. Although the same behaviour 
is observed in all the environments probed, the relative fractions strongly 
depend on the environment and the morphological type. A different trend is observed 
in the GV where, with the exception of field galaxies, the fraction of galaxies 
remains nearly constant with $M_{\ast}$.

The fraction of galaxies in the PS is shown in the top row of Fig. 
\ref{fig:frac}, where we can see that the highest values are found in the
inner region of X-ray clusters. Galaxies at the outskirts of clusters and 
in groups have very similar fractions, not only in the PS, but also in the GV 
and the SFS. The lowest relative fraction of galaxies in the PS corresponds to 
the field, and is nearly zero at the lowest mass bin. For this bin, a low value 
is also observed for field galaxies in the GV, indicating that a higher density
environment is necessary in order to transform galaxies with
$M_{\ast}\sim 10^{10} M_{\odot}$. It can also be seen in Fig. \ref{fig:frac}  
that the fraction of high mass galaxies in the PS is extremely high and similar 
in galaxy systems (XC$_{\rm in}$, XC$_{\rm out}$, and groups). This 
result suggests that massive galaxies do not need the extreme conditions of
cluster cores to complete their transformation from the SFS to the PS.
In the bottom row of Fig. \ref{fig:frac} we show the relative fraction of 
SFS galaxies as a function of stellar mass. The abundance of these galaxies 
also depends on the environment and, again, XC$_{\rm out}$ and groups have similar 
values. 

In the central row of this figure we show the fraction of GV galaxies 
as a function of $M_{\ast}$. With the exception of field galaxies, the GV 
is populated by objects that show a weak dependence on stellar mass, environment
and morphological type. On average, the fraction of galaxies in the GV is 
$\sim20\%$. It is worth noticing that this is also the case for galaxies 
in the cluster core, where the reservoir of galaxies to be quenched is quite 
small. The lack of a strong dependence of the fraction of galaxies in the GV
with $M_{\ast}$, could be the result of two opposite effects: an increase 
in the likelihood and/or relative strength of quenching processes as $M_{\ast}$ 
grows, and a decrement with $M_{\ast}$ of the number of galaxies in the SFS to be 
quenched. The population of field galaxies in the GV grows, with the stellar 
mass being nearly zero for the lower bin of $M_{*}$, similar to what was 
observed for galaxies in the PS.  

\subsection{The specific star formation rate as a function of stellar mass}

\begin{figure*}
\includegraphics[width=15cm]{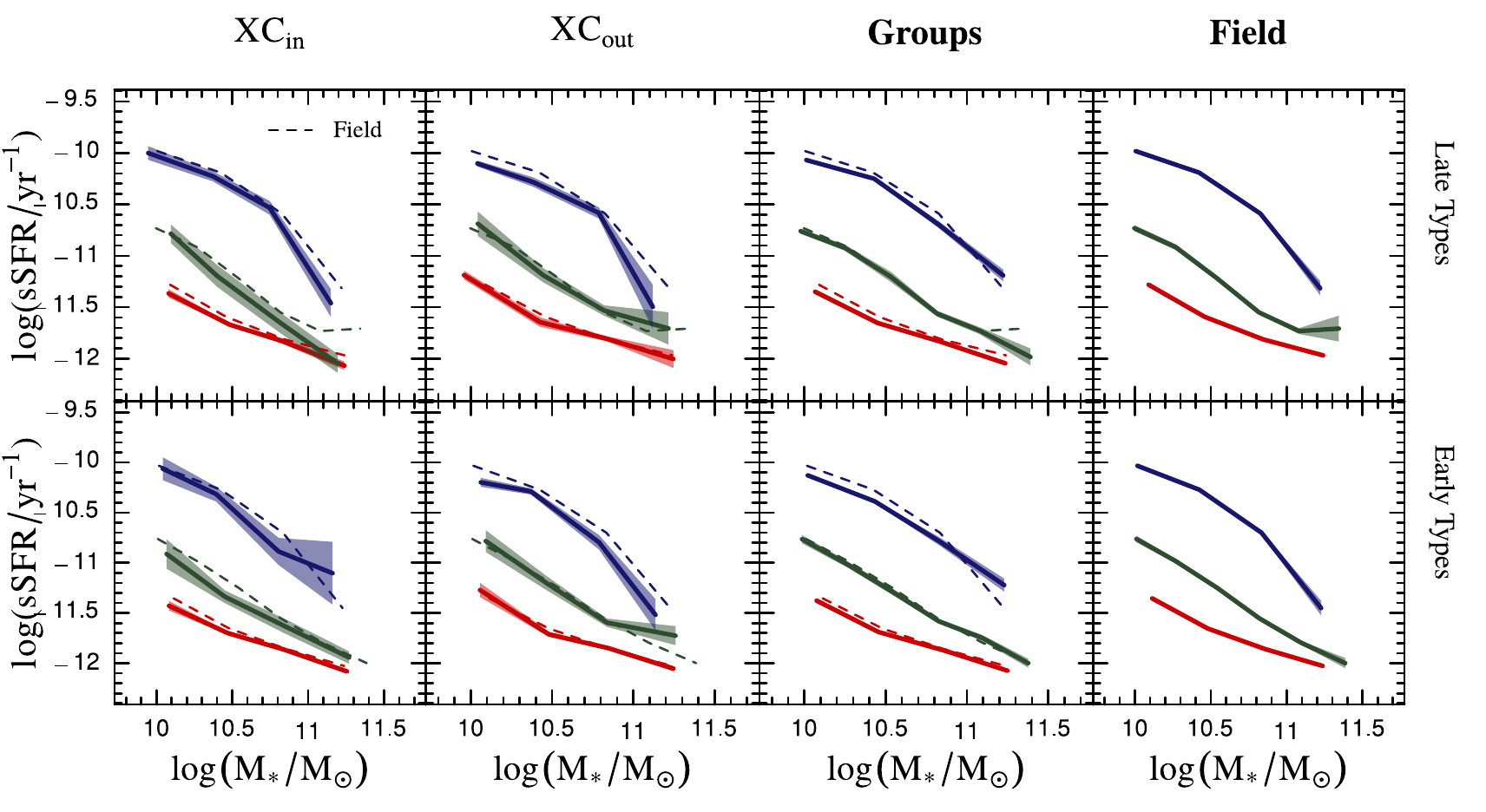}
\caption{The specific star formation rate (sSFR) as a function of stellar mass 
for each environment. {\em Top panels} correspond to late-type galaxies, and 
{\em bottom panels} to early-type galaxies. 
For a reference, we include the field values as dashed lines in the first three 
columns. Colours and lines are as in Fig. 
\ref{fig:ba}. Error bars were obtained by the bootstrap resampling technique.}
\label{fig:ssfrZoo}
\end{figure*}

In Fig. \ref{fig:ssfrZoo}, we show the median specific star formation rate, 
as a function of the 
stellar mass for PS, GV and SFS galaxies. In the 
case of star forming galaxies, this is known as the \textit{main sequence}
\citep{Brinchmann:2004}. To facilitate the comparison, in this figure,
we include, in the denser environment panels, the trends of 
galaxies in the field as dashed lines. Clearly, sSFR anti-correlates with $M_{\ast}$, 
which may be a consequence of the well-known increase in bulge mass-fractions (i.e. 
portions of a galaxy that do not form stars) with $M_{\ast}$ 
\citep{Abramson:2014}. Fig. \ref{fig:ssfrZoo} shows that the sSFR-$M_{\ast}$ 
relation in the GV is far from the \textit{main sequence}.
Both late- and early-type galaxies in the GV are also off the main sequence.
The PS of galaxies lies further off the main sequence than green galaxies. 
We find no evidence of an environmental dependence of the sSFR-stellar
mass relation.

\subsection{The history of star formation in the inner regions of galaxies}

The VESPA catalogue \citep{Tojeiro:2007, Tojeiro:2009} provides a high resolution
star formation history (SFH), which covers the entire age of the Universe by 
means of 16 time ages, equally spaced in logarithm of time. 
VESPA SFHs are derived from SDSS spectroscopy, therefore they are a measure
of the SFH in the inner parts of galaxies. Therefore, the results of this section 
concern only central processes. The authors of VESPA warn that their high 
resolution SFHs should be used carefully, because they are not generally 
reliable in a single object basis \citep{Tojeiro:2009}. 
However, these SFHs can be used to perform statistical studies involving 
several individual objects. We have chosen VESPA's {\tt runID1} model for our 
computations. This model uses \citet{Bruzual:2003} spectral synthesis 
populations and considers a one-parameter dust model, which does not 
consider extra extinction for younger stellar populations. 
Since we are interested in studying the overall history of stellar build-up in 
our galaxies, adopting a two-dust model that gives a 
special treatment to young stellar population would not result in an 
improvement of the resulting SFHs. 
We have checked that there are no qualitative variations in the results that 
follow if we use either two-dust models or \citet{Maraston:2005} 
spectral synthesis population models
(i.e. VESPA's {\tt runID2}, {\tt runID3} and {\tt runID4} models).

\begin{figure*}
\includegraphics[width=15cm]{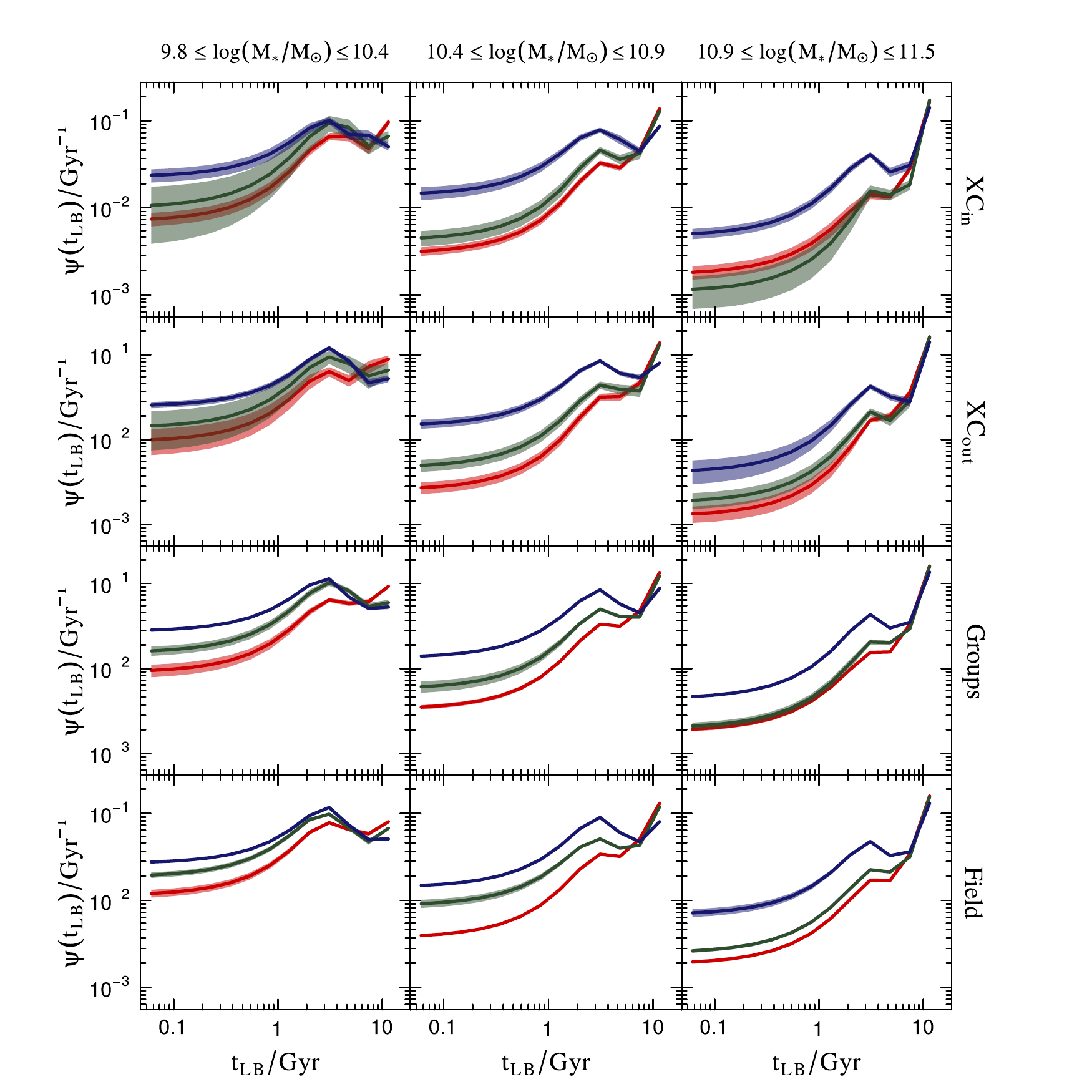}
\caption{The SFR normalised to the total mass formed throughout the history of the Universe as a function of the lookback time (see Eq. \ref{eq:sfr}) for a stacking
of galaxies as described in the text.
All functions shown have been convoluted with a Gaussian kernel.
Columns correspond to different stellar masses, as quoted at the top.
Rows correspond to different environments, as quoted on the right.
Error-bars were computed using the bootstrap resampling technique.
Colours are as in Fig. \ref{fig:ba}}
\label{fig:sfr}
\end{figure*}

In this part of our analysis, for each sequence and environment, we stack 
the galaxies into 3 bins in stellar mass, equally spaced in logarithm of mass, 
in the range $9.8\le\log(M/M_{\odot})\le11.5$, and compute an averaged SFH. 
We basically construct the average SFH of all the stellar mass formed by the 
galaxies in the bin. Let us consider the $k-$th bin, which contains $N_k$ 
galaxies with stellar masses $M_i$ in the range 
$M_{\rm min}^{(k)}< M_i\le M_{\rm max}^{(k)}$. 
Let $m_i(t_j)$ be the stellar mass formed by the galaxy $i$ in the age $t_j$, 
thus the total amount of stellar mass formed by this galaxy throughout its 
lifetime will be\footnote{VESPA models label the most recent age as 0 
and the oldest as 15} 
$M_i^T=\sum_{j=0}^{15}m_i(t_j)$. 
The fraction of stellar mass formed in the age $j$, relative
to the total mass amount of mass formed at all ages, by all galaxies stacked 
in the bin is:
\begin{equation}
f_k(t_j)=\frac{\sum_{i=1}^{N_k} m_i(t_j)}{\sum_{i=1}^{N_k} M_i^T}.
\label{eq:frac}
\end{equation}
If we divide $f_k(t_j)$ by the time width of the $j-$th age,
$\Delta t_j$, we obtain a star formation rate for that age (note that VESPA's high 
resolution models are computed assuming that the SFRs of galaxies 
are constant within each age), normalised to the total amount of mass formed by the 
galaxies in the $k-$th bin during their lifetimes:
\begin{equation}
\psi_k(t_j)=\frac{\sum_{i=1}^{N_k} m_i(t_j)}
{\Delta t_j\sum_{i=1}^{N_k} M_i^T}.
\label{eq:sfr}
\end{equation}
We show in Fig. \ref{fig:sfr} the normalised SFR as a function of the look-back time 
($t_{LB}$), for each mass bin, environment and sequence. These SFRs were 
convoluted with a Gaussian kernel with $\sigma=500\Myr$.
We assume all galaxies started their star formation process at $t_{LB}=14~{\rm Gyr}$.
Clearly, the star formation history depends on stellar mass, sequence and 
environment. In general we observe that, in the last $\sim4~{\rm Gyr}$,
as was expected, SFS galaxies have been the most actively star-forming,
followed in decreasing mode by GV and PS.  
Another overall conclusion from Fig. \ref{fig:sfr} is that the more massive 
the galaxies, the earlier they formed most of their stellar mass. An indication
of effects of the environment in the history of star formation in galaxies 
is evident for GV galaxies: they have SFRs that become more similar to those of
PS galaxies as we move from the field to higher density environments.

\begin{figure}
\includegraphics[width=9cm]{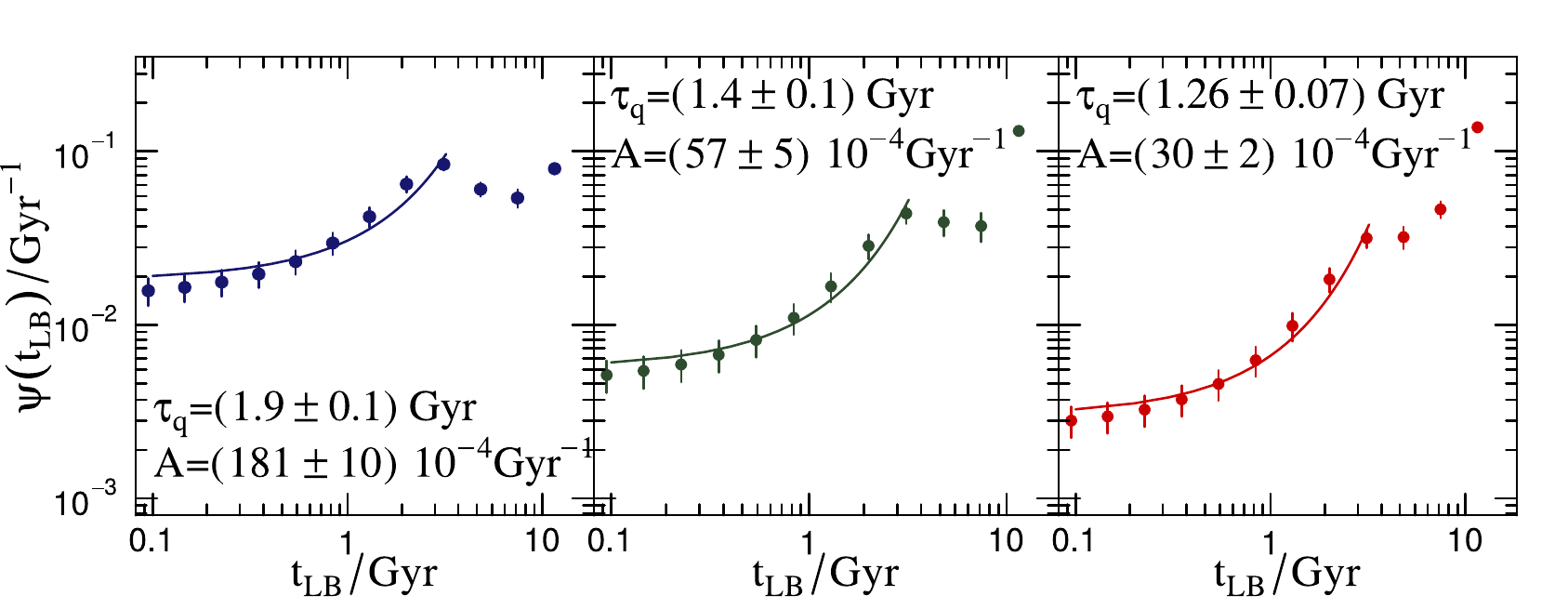}
\caption{The normalised SFR as a function of the 
lookback time for intermediate mass galaxies at the outskirts of
X-ray clusters (second column, second row of Fig. \ref{fig:sfr}).
{\em Left panel:} SFS. {\em Central panel:} GV. {\em Right panel:} PS.
Solid lines are the best-fitting function of the form of Eq. \ref{eq:fit}. 
Inside each panel we quote the quenching time $\tau_q$ and the amplitude $A$.
}
\label{fig:fit}
\end{figure}
 
The normalised star formation rates shown in Fig. \ref{fig:sfr} 
rapidly decay with cosmic time during the last $\sim 4~\Gyr$, and
can be well described by a decaying (growing) exponential function of cosmic (lookback) 
time from $t_{LB}\sim 4~\Gyr$ to the present:
\begin{equation}
\psi(t_{LB})= A~\exp\left(\frac{t_{LB}}{\tau_q}\right).
\label{eq:fit}
\end{equation}
We associate the characteristic time-scale, $\tau_q$, to a quenching time. 
We show in Fig. \ref{fig:fit} examples of these  
normalised SFR and the best-fitting exponential models (Eq. \ref{eq:fit}). 
Best-fitting parameters for all environments, sequences and mass bins
are quoted in Table \ref{tab:fit}.

\begin{figure*}
\includegraphics[width=15cm]{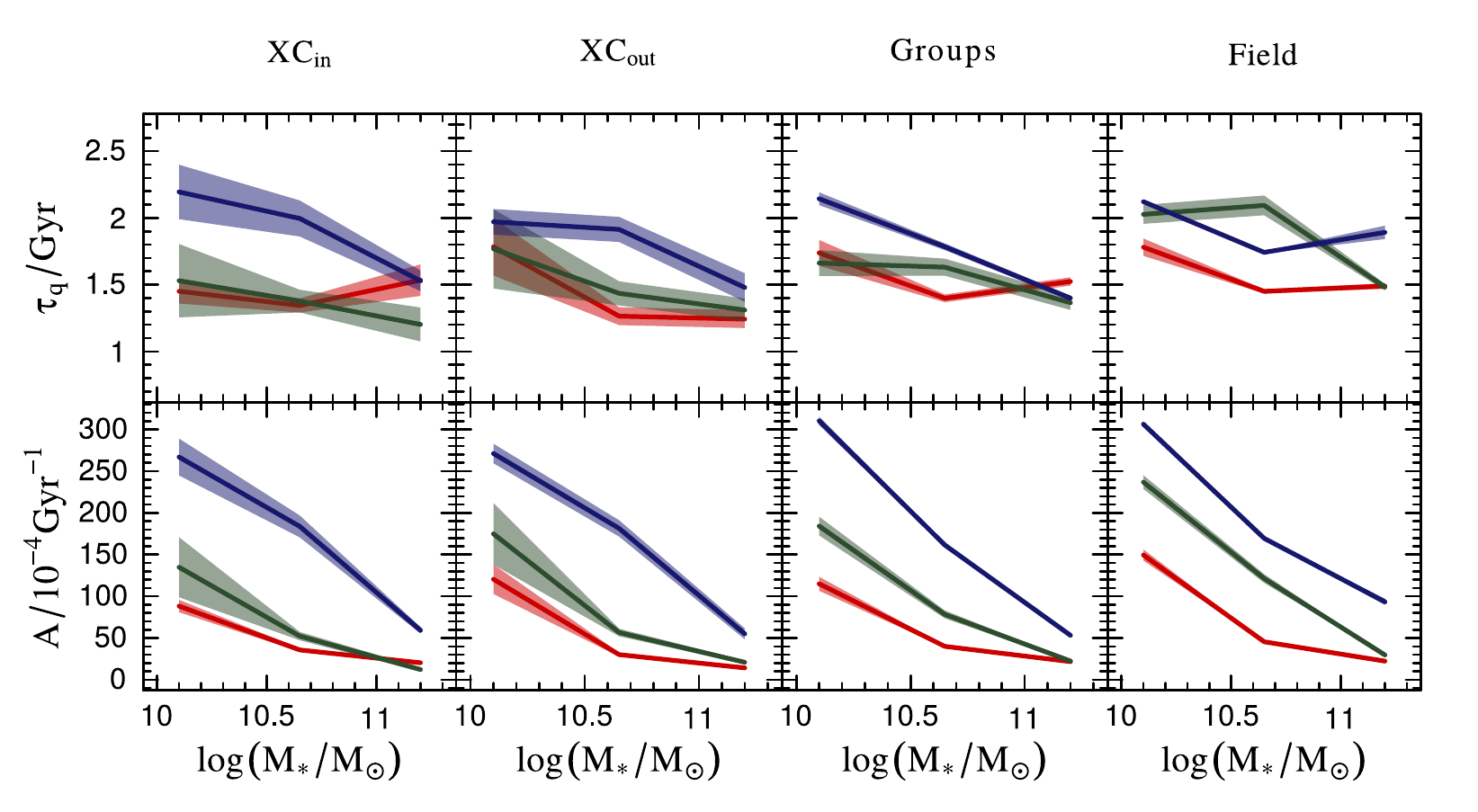}
\caption{
The stellar mass dependence of the best-fitting quenching time $\tau_q$ 
(upper panels) and amplitude $A$ (lower panels) to the last
$4~\Gyr$ of the SFRs in Fig. \ref{fig:sfr}.
Panels correspond to different environments, from left to right: 
inner regions of X-ray clusters, outer region of X-ray clusters, groups,
and field. Colours are as in Fig. \ref{fig:ba}.}
\label{fig:tau_mass}
\end{figure*}

\begin{figure*}
\includegraphics[width=15cm]{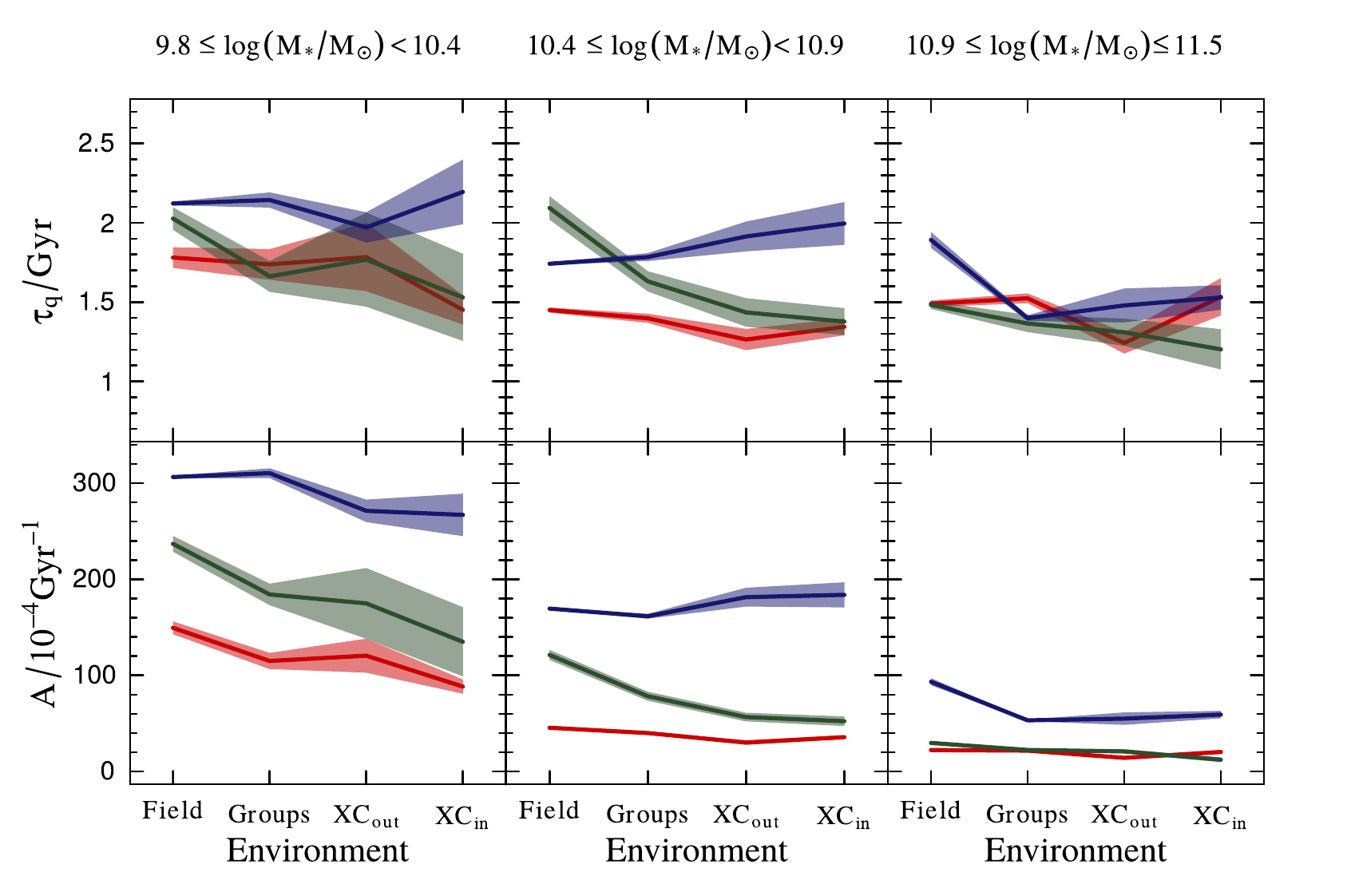}
\caption{
The environment dependence of the best-fitting quenching time $\tau_q$ 
(upper panels) and amplitude $A$ (lower panels) to the last
$4~\Gyr$ of the SFRs in Fig. \ref{fig:sfr}.
Panels correspond to different stellar mass bins, as quoted at the top.
Colours are as in Fig. \ref{fig:ba}.}
\label{fig:tau_env}
\end{figure*}

We show in Fig. \ref{fig:tau_mass}, for each environment and sequence, the 
quenching time and the amplitude of the normalised SFR in the last $4~\Gyr$ as a 
function of the stellar mass. In general, quenching times are longer for
SFS galaxies and shorter for PS galaxies. With the exception of PS galaxies in the inner 
regions of X-ray clusters, all sequences show quenching times that decrease 
with increasing mass. It is worth noticing that the quenching times of the PS and 
the GV of galaxies in clusters are very similar.
Regarding the amplitude of the normalised SFR, the largest values correspond to SFS,
intermediate values to GV and the smallest values to PS.
All sequences have a decreasing trend with mass. This effect is stronger 
for the SFS. An environmental dependence of the amplitude of the SFR for 
GV galaxies is seen in this figure: moving from the field to the inner regions
of clusters, it becomes more similar to the PS values.

The effects of the environment on star formation are more clearly seen in 
Fig. \ref{fig:tau_env}, where we plot the quenching time and the amplitude of 
the normalised SFRs as a function of the environment. 
This figure shows that the GV is the sequence that has the most consistent 
trend with the environment: regardless the mass bin, both the quenching time
and the amplitude decay monotonically with increasing density environment.
This consistency is not seen for SFS and PS.

\begin{table*}
\caption{Best-fitting parameters of the form in Eq. \ref{eq:fit} to the last
$4~\Gyr$ of the SFRs in Fig. \ref{fig:sfr}.}
\begin{tabular}[ht]{cccccccc}
\hline
&& \multicolumn{2}{c}{Star Forming Sequence}   &
   \multicolumn{2}{c}{Green Valley} &
   \multicolumn{2}{c}{Passive Sequence} \\
\hline
Environment &Stellar  & $A$ & $\tau_q$ & 
                 $A$ & $\tau_q$ &
                 $A$ & $\tau_q$ \\
&mass bin   & [$10^{-4}{\rm Gyr}^{-1}$] & [Gyr] &
              [$10^{-4}{\rm Gyr}^{-1}$] & [Gyr] &
              [$10^{-4}{\rm Gyr}^{-1}$] & [Gyr] \\

\hline
Field      
& 1 & $306\pm2$ & $2.12\pm0.01$ & $237\pm8$ & $2.03\pm0.07$ & $150\pm7$ & $1.78\pm0.06$\\
& 2 & $170\pm1$ & $1.74\pm0.01$ & $121\pm5$ & $2.09\pm0.07$ & $46\pm1$ & $1.45\pm0.02$\\
& 3 & $93\pm4$ & $1.89\pm0.05$ & $30\pm1$ & $1.48\pm0.02$ & $ 22\pm1$ & $1.49\pm0.02$\\
\hline
Groups    
& 1 & $310\pm5$ & $2.14\pm0.05$ & $184\pm11$ & $1.7\pm0.1$ & $115\pm8$ & $1.7\pm0.1$\\
& 2 & $162\pm3$ & $1.78\pm0.02$ & $78\pm5$ & $1.63\pm0.06$ & $40\pm1$ & $1.40\pm0.03$\\
& 3 & $53\pm1$ & $1.40\pm0.02$ & $22\pm1$ & $1.36\pm0.05$ & $22\pm1$ & $1.52\pm0.03$\\
\hline
XC$_{\rm out}$ 
& 1 & $271\pm12$ & $2.0\pm0.1$ & $175\pm37$ & $1.8\pm0.3$ & $121\pm18$ & $1.8\pm0.2$\\
& 2 & $181\pm10$ & $1.9\pm0.1$ & $57\pm5$ & $1.4\pm0.1$ & $30\pm2$ & $1.26\pm0.07$\\
& 3 & $55\pm7$ & $1.5\pm0.1$ & $ 21\pm2$ & $1.3\pm0.1$ & $ 14\pm2$ & $1.24\pm0.07$\\
\hline
XC$_{\rm in}$  
& 1 & $267\pm22$ & $2.2\pm0.2$ & $135\pm36$ & $1.5\pm0.3$ & $88\pm7$ & $1.5\pm0.1$\\
& 2 & $184\pm13$ & $2.0\pm0.1$ & $52\pm5$ & $1.4\pm0.1$ & $36\pm2$ & $1.34\pm0.05$\\
& 3 & $59\pm4$ & $1.5\pm0.1$ & $12\pm2$ & $1.2\pm0.1$ & $20\pm2$ & $1.5\pm0.1$\\
\hline
\label{tab:fit}
\end{tabular}
\end{table*}

From Figs. \ref{fig:tau_mass} and \ref{fig:tau_env} we were able to conclude that the
quenching of the star formation in SFS galaxies is primarily determined by their mass, 
whereas for GV galaxies it is more sensitive to the environment. The 
quenching time of PS galaxies appears to be less dependent on mass and 
environment.   


\section{Conclusions and discussion}\label{conclu}

To shed light on the impact of internal and external quenching mechanisms
upon galaxies, in this paper we compare properties of star-forming, passive 
and transition galaxies in four discrete environments: field; groups as 
representative of intermediate mass systems; and the most massive virialised 
systems in the Universe, X-ray clusters, distinguishing between their
inner and outer regions.
We construct samples of galaxies in these environments that are bound to have
similar redshift distributions out of the SDSS. We classify galaxies into 
three sequences: passive, green valley and star-forming, by means of their 
UV-optical colour $^{0.1}(NUV-r)$. 
We study a number of galaxy properties: stellar mass, 
morphology, specific star formation rate and the history of star formation.\\

Our main findings can be summarised as:
\begin{enumerate}
\item Regarding the morphological classification, as is well established 
in the literature, GV galaxies have intermediate morphologies. We show that 
this appears to be independent of the environment, with the possible
exception of the inner regions of X-ray clusters where few green spirals are found.

\item When comparing GV stellar masses across environments, 
field galaxies tend to be more 
massive than in other environments. The environment that has the 
largest fraction of low mass GV galaxies is the outskirts of clusters.
If internal quenching is more efficient in massive galaxies, the
environment should play a central role in quenching lesser massive galaxies, as 
our results suggest.

\item In contrast to the growing (decaying) trend of the abundance of PS (SFS) 
galaxies as a function of stellar mass, seen in all environments, the 
abundance of GV galaxies is almost constant, with the exception of 
the field. On average, GV galaxies account for $\sim20\%$ of all galaxies
in groups and X-ray clusters. The field differs from the other environments
in that it has a clear lack of $\sim 10^{10}M_{\odot}$ GV and PS galaxies.
This is another indication that high density environments are needed to 
transform lower mass galaxies. At the high mass end, we observe that the fraction
of galaxies in the PS is similar for the three densest environments, suggesting
that, beyond a certain (high) density threshold, efficiency in quenching
galaxies is independent of environment. Similarly, \citet{Martinez:2008}
suggested that, above a certain cluster mass, galaxies in clusters experience the
same physical processes acting with similar relative effectiveness,
thus producing a saturation in the mass-colour relationship.

\item When analysing the relationship between the specific star formation 
rate and the stellar mass, we find that GV galaxies lie far off the main 
sequence, and closer to PS galaxies. We find that the sSFR-stellar mass relation
does not depend on environment.

\item Using VESPA data, we find that the stacked star formation history in the
inner regions of galaxies depends on sequence and environment. 
In general, over the last $\sim 4 \Gyr$, GV galaxies have star formation rates 
intermediate between SFS and PS galaxies. As denser environments are considered, 
the history of star formation of GV galaxies becomes more similar to that 
of PS galaxies.

\item Using a simple decaying exponential model to describe the star formation 
rate of galaxies in the last $4~\Gyr$, we estimate the quenching time 
and amplitude of the SFR as a function of sequence and
environment. As expected, the longest quenching times 
and largest amplitudes correspond to SFS. PS galaxies 
have typically the shortest quenching times and smallest amplitudes. 
In most cases, GV galaxies have intermediate values.
The GV is the sequence that has the clearest and most consistent dependence on 
the environment: both the quenching time and the amplitude decrease with 
increasing environment density.
\end{enumerate}

Our results indicate that external quenching sources, have an important 
role in galaxy evolution. The relative impact upon galaxies of internal and 
external quenching processes clearly depends on environment.
Since field galaxies are less affected by environmental processes, internal 
quenching should be their main driver of transformation. 
As we move from the field to denser environments, the physical mechanisms
responsible for external quenching become more efficient, as can be seen from
the higher fraction of PS galaxies in these environments.
Quenching times in these denser environments are shorter than in the field,
and the physical processes that determine external quenching can account for
these differences. The impact of dense environments, such as groups, has 
been reported by \citet{Rasmussen:2012}, favouring average quenching timescales 
of $\gtrsim 2\Gyr$ and suggesting the simultaneous action of tidal interactions
and starvation. These timescales are close to our findings 
and larger than those $<1~\Gyr$ found by \citet{Crossett:2016}.

The fact that the fraction of transition galaxies remains roughly
constant with stellar mass in dense environments may be explained
in term of two opposite effects: an increase 
in the likelihood and/or relative strength of quenching processes as $M_{\ast}$ 
increases, and a decrement with $M_{\ast}$ of the number of galaxies in the SFS to be 
quenched.
 
\section*{Acknowledgements}
We thank the anonymous referee for comments and suggestions that improved
this paper. 

This paper was partially supported by CONICET grants PIP 11220130100365CO
and 11220120100492CO, and grants from SeCyT, Universidad Nacional de 
C\'ordoba, Argentina.

Funding for the Sloan Digital Sky Survey IV has been provided by
the Alfred P. Sloan Foundation, the U.S. Department of Energy Office of
Science, and the Participating Institutions. SDSS-IV acknowledges
support and resources from the Center for High-Performance Computing at
the University of Utah. The SDSS web site is www.sdss.org.
SDSS-IV is managed by the Astrophysical Research Consortium for the 
Participating Institutions of the SDSS Collaboration including the 
Brazilian Participation Group, the Carnegie Institution for Science, 
Carnegie Mellon University, the Chilean Participation Group, the French 
Participation Group, Harvard-Smithsonian Center for Astrophysics, 
Instituto de Astrof\'isica de Canarias, The Johns Hopkins University, 
Kavli Institute for the Physics and Mathematics of the Universe (IPMU) / 
University of Tokyo, Lawrence Berkeley National Laboratory, 
Leibniz Institut f\"ur Astrophysik Potsdam (AIP),  
Max-Planck-Institut f\"ur Astronomie (MPIA Heidelberg), 
Max-Planck-Institut f\"ur Astrophysik (MPA Garching), 
Max-Planck-Institut f\"ur Extraterrestrische Physik (MPE), 
National Astronomical Observatories of China, New Mexico State University, 
New York University, University of Notre Dame, 
Observat\'ario Nacional / MCTI, The Ohio State University, 
Pennsylvania State University, Shanghai Astronomical Observatory, 
United Kingdom Participation Group,
Universidad Nacional Aut\'onoma de M\'exico, University of Arizona, 
University of Colorado Boulder, University of Oxford, University of Portsmouth, 
University of Utah, University of Virginia, University of Washington, University of Wisconsin, 
Vanderbilt University, and Yale University.

Galaxy Evolution Explorer (GALEX) is a NASA Small Explorer,
launched in 2003. We acknowledge NASA’s support for construction, operation 
and science analysis for the GALEX mission, developed in cooperation with 
the Centre National
d’Etudes Spatiales of France and the Korean Ministry of Science and
Technology.

\end{document}